
 \catcode`@=11
 \newskip\ttglue

 \font\twelverm=cmr12
 \font\eightrm=cmr8   \font\sixrm=cmr6
 \font\eighti=cmmi8   \font\sixi=cmmi6
 \font\eightsy=cmsy8  \font\sixsy=cmsy6
 \font\eightbf=cmbx8  \font\sixbf=cmbx6
 \font\eighttt=cmtt8
 \font\eightit=cmti8
 \font\eightsl=cmsl8

 \def\rm{\fam0\tenrm}%
 \textfont0=\tenrm \scriptfont0=\sevenrm \scriptscriptfont0=\fiverm
 \textfont1=\teni  \scriptfont1=\seveni  \scriptscriptfont1=\fivei
 \textfont2=\tensy \scriptfont2=\sevensy \scriptscriptfont2=\fivesy
 \textfont3=\tenex \scriptfont3=\tenex   \scriptscriptfont3=\tenex
 \textfont\itfam=\tenit	\def\it{\fam\itfam\tenit}
 \textfont\slfam=\tensl \def\sl{\fam\slfam\tensl}
 \textfont\ttfam=\tentt \def\tt{\fam\ttfam\tentt}
 \textfont\bffam=\tenbf \def\bf{\fam\bffam\tenbf}
 \scriptfont\bffam\sevenbf \scriptscriptfont\bffam=\fivebf
 \tt \ttglue=.5em plus.25em minus.15em
 \normalbaselineskip=12pt
 \setbox\strutbox=\hbox{\vrule height8.5pt depth3.5pt width0pt}
 \let\sc=\eightrm \normalbaselines\rm

\def\eightpoint{\def\rm{\fam0\eightrm}
 \textfont0=\eightrm \scriptfont0=\sixrm \scriptscriptfont0=\fiverm
 \textfont1=\eighti  \scriptfont1=\sixi  \scriptscriptfont1=\fivei
 \textfont2=\eightsy \scriptfont2=\sixsy \scriptscriptfont2=\fivesy
 \textfont3=\tenex   \scriptfont3=\tenex \scriptscriptfont3=\tenex
 \textfont\itfam=\eightit  \def\it{\fam\itfam\eightit}
 \textfont\slfam=\eightsl  \def\sl{\fam\slfam\eightsl}
 \textfont\ttfam=\eighttt  \def\tt{\fam\ttfam\eighttt}
 \textfont\bffam=\eightbf  \scriptfont\bffam\sixbf
 \scriptscriptfont\bffam=\fivebf  \def\bf{\fam\bffam\eightbf}
 \tt \ttglue=.5em plus.25em minus.15em
 \normalbaselineskip=11pt
 \setbox\strutbox=\hbox{\vrule height7pt depth2pt width0pt}
 \let\sc=\sixrm \normalbaselines\rm}

 \def\simlt{\mathrel{\hbox to 0pt{\lower 3.5pt\hbox{$\mathchar"218$}\hss}
   \raise 1.5pt\hbox{$\mathchar"13C$}}}
 \def\simgt{\mathrel{\hbox to 0pt{\lower 3.5pt\hbox{$\mathchar"218$}\hss}
   \raise 1.5pt\hbox{$\mathchar"13E$}}}
 \def\ref{\par\noindent\parskip=0pt\hangindent12pt\hangafter1}

 \vsize=53pc
 \hsize=7.in
 \hoffset=-0.2truein
 \parskip=0pt
 \parindent=12pt
 \baselineskip=12pt
 \lineskip=-12pt
 \lineskiplimit=-24pt
 \pretolerance=100
 \tolerance=1000
 \relpenalty=50
 \vbadness=10000
 \hbadness=10000
 \predisplaypenalty=50
 \postdisplaypenalty=50 
 \topskip=12pt
 \splittopskip=\topskip
 \maxdepth=4pt
 \splitmaxdepth=\maxdepth

 \abovedisplayskip=12pt
 \belowdisplayskip=12pt
 \abovedisplayshortskip=0pt
 \belowdisplayshortskip=12pt

 \def\journal#1{\null\vskip-12pt
   \noindent {\eightrm T{\sixrm HE} A{\sixrm STROPHYSICAL} J{\sixrm OURNAL}
   {\eightrm #1}}
   \vskip-12pt}
 \def\title#1{\null
   \vskip48pt
   \centerline{\twelverm #1}
   \vskip12pt}
 \def\titlem#1{\vskip-6pt
   \centerline{\twelverm #1}
   \vskip6pt}
 \def\author#1{
   \centerline{#1}}
 
 \def\date#1#2{
   \centerline {\eightsl Received #1; accepted #2}}
 \def\abstract{
   \vskip18pt
   \centerline{ABSTRACT}
   \vskip10pt
   \par\parshape=1 0.5truein 6.0truein}
 
 \def\subject{
   \vskip2pt
   \par\parshape=1 0.5truein 6.0truein
   \noindent
   {\it Subject headings: } }
 \def\maintext{
   \vskip24pt\begindoublecolumns\vskip-12pt}
 \def\section#1{
   \vskip18pt
   \centerline{#1}
   \vskip6pt}
 \def\subsection#1{
   \vskip10pt
   \centerline{#1}
   \vskip2pt}

\def\fdg{\hbox{$.\!\!^\circ$}}
\def\arcmin{\hbox{$^\prime$}}
\def\micron{\hbox{$\mu$m}}
\def\arcdeg{\hbox{$^\circ$}}
\def\lesssim{\mathrel{\hbox{\rlap{\hbox{\lower4pt\hbox{$\sim$}}}\hbox{$<$}}}}
\def\gtrsim{\mathrel{\hbox{\rlap{\hbox{\lower4pt\hbox{$\sim$}}}\hbox{$>$}}}}

 \newdimen\h@size
 \h@size=\hsize

 \def\figtop#1#2#3{\topinsert\null\vskip-12pt
  \psfig{file={#1},width=\h@size,silent=yes}
  \vskip-6pt
  \edef\h@old{\the\hsize}\hsize=\h@size
  Fig.~{#2}.---{#3}\vskip12pt\vskip-5.23433pt
  \hsize=\h@old\endinsert\vfuzz=0.5pt}
 
 \def\tabbot#1#2{\null\vskip-24pt
   \edef\h@old{\the\hsize}\hsize=\h@size
   \vskip15pt
   \centerline {{\tenrm TABLE} {\tenrm{#1}}}\vskip4pt
   \centerline {\tenrm{#2}}
   \tabskip=0.5em plus 10.0em minus 0.5em
   \halign to \hsize}
 \def\tabhead#1{
   \noalign{\vskip10pt\hrule\vskip2pt\hrule\vskip6pt}#1}
 \def\tabbody#1{\noalign{\vskip6pt\hrule\vskip6pt}
   #1\noalign{\vskip6pt\hrule\vskip8pt}}
 \def\tabnote#1{\noindent{#1}\vskip17pt
   \vskip-36pt\hsize=\h@old\vfuzz=0.5pt}

 \def\footnoterule{\null}

%
%
\newcount\Chno\newcount\Eqno\newcount\Eqlt
\def\nxtChno{\global\advance\Chno by 1 \global\Eqno=0 \global\Eqlt=`a}
\def\nxtEqno{\global\advance\Eqno by 1 \global\Eqlt=`a}
\def\nxtEqlt{\global\advance\Eqlt by 1}
\def\ChEq{\the\Eqno}
\def\ChEql{\ChEq\char\the\Eqlt}

%
%
\def\PsfigVersion{1.9}
\ifx\undefined\psfig\else\endinput\fi

%

\let\LaTeXAtSign=\@
\let\@=\relax
\edef\psfigRestoreAt{\catcode`\@=\number\catcode`@\relax}
\catcode`\@=11\relax
\newwrite\@unused
\def\ps@typeout#1{{\let\protect\string\immediate\write\@unused{#1}}}
\ps@typeout{psfig/tex \PsfigVersion}


\def\figurepath{./}

%
%
\def\@nnil{\@nil}
\def\@empty{}
\def\@psdonoop#1\@@#2#3{}
\def\@psdo#1:=#2\do#3{\edef\@psdotmp{#2}\ifx\@psdotmp\@empty \else
    \expandafter\@psdoloop#2,\@nil,\@nil\@@#1{#3}\fi}
\def\@psdoloop#1,#2,#3\@@#4#5{\def#4{#1}\ifx #4\@nnil \else
       #5\def#4{#2}\ifx #4\@nnil \else#5\@ipsdoloop #3\@@#4{#5}\fi\fi}
\def\@ipsdoloop#1,#2\@@#3#4{\def#3{#1}\ifx #3\@nnil 
       \let\@nextwhile=\@psdonoop \else
      #4\relax\let\@nextwhile=\@ipsdoloop\fi\@nextwhile#2\@@#3{#4}}
\def\@tpsdo#1:=#2\do#3{\xdef\@psdotmp{#2}\ifx\@psdotmp\@empty \else
    \@tpsdoloop#2\@nil\@nil\@@#1{#3}\fi}
\def\@tpsdoloop#1#2\@@#3#4{\def#3{#1}\ifx #3\@nnil 
       \let\@nextwhile=\@psdonoop \else
      #4\relax\let\@nextwhile=\@tpsdoloop\fi\@nextwhile#2\@@#3{#4}}
%
\ifx\undefined\fbox
\newdimen\fboxrule
\newdimen\fboxsep
\newdimen\ps@tempdima
\newbox\ps@tempboxa
\fboxsep = 3pt
\fboxrule = .4pt
\long\def\fbox#1{\leavevmode\setbox\ps@tempboxa\hbox{#1}\ps@tempdima\fboxrule
    \advance\ps@tempdima \fboxsep \advance\ps@tempdima \dp\ps@tempboxa
   \hbox{\lower \ps@tempdima\hbox
  {\vbox{\hrule height \fboxrule
          \hbox{\vrule width \fboxrule \hskip\fboxsep
          \vbox{\vskip\fboxsep \box\ps@tempboxa\vskip\fboxsep}\hskip 
                 \fboxsep\vrule width \fboxrule}
                 \hrule height \fboxrule}}}}
\fi
%
%
\newread\ps@stream
\newif\ifnot@eof       
\newif\if@noisy        
\newif\if@atend        
\newif\if@psfile       
%
%
{\catcode`\%=12\global\gdef\epsf@start{
\def\epsf@PS{PS}
\def\epsf@getbb#1{%
%
%
\openin\ps@stream=#1
\ifeof\ps@stream\ps@typeout{Error, File #1 not found}\else
%
%
   {\not@eoftrue \chardef\other=12
    \def\do##1{\catcode`##1=\other}\dospecials \catcode`\ =10
    \loop
       \if@psfile
	  \read\ps@stream to \epsf@fileline
       \else{
	  \obeyspaces
          \read\ps@stream to \epsf@tmp\global\let\epsf@fileline\epsf@tmp}
       \fi
       \ifeof\ps@stream\not@eoffalse\else
%
%
       \if@psfile\else
       \expandafter\epsf@test\epsf@fileline:. \\%
       \fi
%
%
          \expandafter\epsf@aux\epsf@fileline:. \\%
       \fi
   \ifnot@eof\repeat
   }\closein\ps@stream\fi}%
%
%
\long\def\epsf@test#1#2#3:#4\\{\def\epsf@testit{#1#2}
			\ifx\epsf@testit\epsf@start\else
\ps@typeout{Warning! File does not start with `\epsf@start'.  It may not be a PostScript file.}
			\fi
			\@psfiletrue} 
%
%
{\catcode`\%=12\global\let\epsf@percent=
%
%
%
\long\def\epsf@aux#1#2:#3\\{\ifx#1\epsf@percent
   \def\epsf@testit{#2}\ifx\epsf@testit\epsf@bblit
	\@atendfalse
        \epsf@atend #3 . \\%
	\if@atend	
	   \if@verbose{
		\ps@typeout{psfig: found `(atend)'; continuing search}
	   }\fi
        \else
        \epsf@grab #3 . . . \\%
        \not@eoffalse
        \global\no@bbfalse
        \fi
   \fi\fi}%
%
%
\def\epsf@grab #1 #2 #3 #4 #5\\{%
   \global\def\epsf@llx{#1}\ifx\epsf@llx\empty
      \epsf@grab #2 #3 #4 #5 .\\\else
   \global\def\epsf@lly{#2}%
   \global\def\epsf@urx{#3}\global\def\epsf@ury{#4}\fi}%
%
%
\def\epsf@atendlit{(atend)} 
\def\epsf@atend #1 #2 #3\\{%
   \def\epsf@tmp{#1}\ifx\epsf@tmp\empty
      \epsf@atend #2 #3 .\\\else
   \ifx\epsf@tmp\epsf@atendlit\@atendtrue\fi\fi}


\chardef\psletter = 11 
\chardef\other = 12

\newif \ifdebug 
\newif\ifc@mpute 
\c@mputetrue 

\let\then = \relax
\def\r@dian{pt }
\let\r@dians = \r@dian
\let\dimensionless@nit = \r@dian
\let\dimensionless@nits = \dimensionless@nit
\def\internal@nit{sp }
\let\internal@nits = \internal@nit
\newif\ifstillc@nverging
\def \Mess@ge #1{\ifdebug \then \message {#1} \fi}

{ 
	\catcode `\@ = \psletter
	\gdef \nodimen {\expandafter \n@dimen \the \dimen}
	\gdef \term #1 #2 #3%
	       {\edef \t@ {\the #1}
		\edef \t@@ {\expandafter \n@dimen \the #2\r@dian}%
		\t@rm {\t@} {\t@@} {#3}%
	       }
	\gdef \t@rm #1 #2 #3%
	       {{%
		\count 0 = 0
		\dimen 0 = 1 \dimensionless@nit
		\dimen 2 = #2\relax
		\Mess@ge {Calculating term #1 of \nodimen 2}%
		\loop
		\ifnum	\count 0 < #1
		\then	\advance \count 0 by 1
			\Mess@ge {Iteration \the \count 0 \space}%
			\Multiply \dimen 0 by {\dimen 2}%
			\Mess@ge {After multiplication, term = \nodimen 0}%
			\Divide \dimen 0 by {\count 0}%
			\Mess@ge {After division, term = \nodimen 0}%
		\repeat
		\Mess@ge {Final value for term #1 of 
				\nodimen 2 \space is \nodimen 0}%
		\xdef \Term {#3 = \nodimen 0 \r@dians}%
		\aftergroup \Term
	       }}
	\catcode `\p = \other
	\catcode `\t = \other
	\gdef \n@dimen #1pt{#1} 
}

\def \Divide #1by #2{\divide #1 by #2} 

\def \Multiply #1by #2
       {{
	\count 0 = #1\relax
	\count 2 = #2\relax
	\count 4 = 65536
	\Mess@ge {Before scaling, count 0 = \the \count 0 \space and
			count 2 = \the \count 2}%
	\ifnum	\count 0 > 32767 
	\then	\divide \count 0 by 4
		\divide \count 4 by 4
	\else	\ifnum	\count 0 < -32767
		\then	\divide \count 0 by 4
			\divide \count 4 by 4
		\else
		\fi
	\fi
	\ifnum	\count 2 > 32767 
	\then	\divide \count 2 by 4
		\divide \count 4 by 4
	\else	\ifnum	\count 2 < -32767
		\then	\divide \count 2 by 4
			\divide \count 4 by 4
		\else
		\fi
	\fi
	\multiply \count 0 by \count 2
	\divide \count 0 by \count 4
	\xdef \product {#1 = \the \count 0 \internal@nits}%
	\aftergroup \product
       }}

\def\r@duce{\ifdim\dimen0 > 90\r@dian \then   
		\multiply\dimen0 by -1
		\advance\dimen0 by 180\r@dian
		\r@duce
	    \else \ifdim\dimen0 < -90\r@dian \then  
		\advance\dimen0 by 360\r@dian
		\r@duce
		\fi
	    \fi}

\def\Sine#1%
       {{%
	\dimen 0 = #1 \r@dian
	\r@duce
	\ifdim\dimen0 = -90\r@dian \then
	   \dimen4 = -1\r@dian
	   \c@mputefalse
	\fi
	\ifdim\dimen0 = 90\r@dian \then
	   \dimen4 = 1\r@dian
	   \c@mputefalse
	\fi
	\ifdim\dimen0 = 0\r@dian \then
	   \dimen4 = 0\r@dian
	   \c@mputefalse
	\fi
	\ifc@mpute \then
		\divide\dimen0 by 180
		\dimen0=3.141592654\dimen0
		\dimen 2 = 3.1415926535897963\r@dian 
		\divide\dimen 2 by 2 
		\Mess@ge {Sin: calculating Sin of \nodimen 0}%
		\count 0 = 1 
		\dimen 2 = 1 \r@dian 
		\dimen 4 = 0 \r@dian 
		\loop
			\ifnum	\dimen 2 = 0 
			\then	\stillc@nvergingfalse 
			\else	\stillc@nvergingtrue
			\fi
			\ifstillc@nverging 
			\then	\term {\count 0} {\dimen 0} {\dimen 2}%
				\advance \count 0 by 2
				\count 2 = \count 0
				\divide \count 2 by 2
				\ifodd	\count 2 
				\then	\advance \dimen 4 by \dimen 2
				\else	\advance \dimen 4 by -\dimen 2
				\fi
		\repeat
	\fi		
			\xdef \sine {\nodimen 4}%
       }}

\def\Cosine#1{\ifx\sine\UnDefined\edef\Savesine{\relax}\else
		             \edef\Savesine{\sine}\fi
	{\dimen0=#1\r@dian\advance\dimen0 by 90\r@dian
	 \Sine{\nodimen 0}
	 \xdef\cosine{\sine}
	 \xdef\sine{\Savesine}}}	      

\def\psdraft{
	\def\@psdraft{0}
}
\def\psfull{
	\def\@psdraft{100}
}

\psfull

\newif\if@scalefirst
\def\psscalefirst{\@scalefirsttrue}
\def\psrotatefirst{\@scalefirstfalse}
\psrotatefirst

\newif\if@draftbox
\def\psnodraftbox{
	\@draftboxfalse
}
\def\psdraftbox{
	\@draftboxtrue
}
\@draftboxtrue

\newif\if@prologfile
\newif\if@postlogfile
\def\pssilent{
	\@noisyfalse
}
\def\psnoisy{
	\@noisytrue
}
\psnoisy
\newif\if@bbllx
\newif\if@bblly
\newif\if@bburx
\newif\if@bbury
\newif\if@height
\newif\if@width
\newif\if@rheight
\newif\if@rwidth
\newif\if@angle
\newif\if@clip
\newif\if@verbose
\def\@p@@sclip#1{\@cliptrue}

\newif\if@decmpr


\def\@p@@sfigure#1{\def\@p@sfile{null}\def\@p@sbbfile{null}
	        \openin1=#1.bb
		\ifeof1\closein1
	        	\openin1=\figurepath#1.bb
			\ifeof1\closein1
			        \openin1=#1
				\ifeof1\closein1%
				       \openin1=\figurepath#1
					\ifeof1
					   \ps@typeout{Error, File #1 not found}
						\if@bbllx\if@bblly
				   		\if@bburx\if@bbury
			      				\def\@p@sfile{#1}%
			      				\def\@p@sbbfile{#1}%
							\@decmprfalse
				  	   	\fi\fi\fi\fi
					\else\closein1
				    		\def\@p@sfile{\figurepath#1}%
				    		\def\@p@sbbfile{\figurepath#1}%
						\@decmprfalse
	                       		\fi%
			 	\else\closein1%
					\def\@p@sfile{#1}
					\def\@p@sbbfile{#1}
					\@decmprfalse
			 	\fi
			\else
				\def\@p@sfile{\figurepath#1}
				\def\@p@sbbfile{\figurepath#1.bb}
				\@decmprtrue
			\fi
		\else
			\def\@p@sfile{#1}
			\def\@p@sbbfile{#1.bb}
			\@decmprtrue
		\fi}

\def\@p@@sfile#1{\@p@@sfigure{#1}}

\def\@p@@sbbllx#1{
		\@bbllxtrue
		\dimen100=#1
		\edef\@p@sbbllx{\number\dimen100}
}
\def\@p@@sbblly#1{
		\@bbllytrue
		\dimen100=#1
		\edef\@p@sbblly{\number\dimen100}
}
\def\@p@@sbburx#1{
		\@bburxtrue
		\dimen100=#1
		\edef\@p@sbburx{\number\dimen100}
}
\def\@p@@sbbury#1{
		\@bburytrue
		\dimen100=#1
		\edef\@p@sbbury{\number\dimen100}
}
\def\@p@@sheight#1{
		\@heighttrue
		\dimen100=#1
   		\edef\@p@sheight{\number\dimen100}
}
\def\@p@@swidth#1{
		\@widthtrue
		\dimen100=#1
		\edef\@p@swidth{\number\dimen100}
}
\def\@p@@srheight#1{
		\@rheighttrue
		\dimen100=#1
		\edef\@p@srheight{\number\dimen100}
}
\def\@p@@srwidth#1{
		\@rwidthtrue
		\dimen100=#1
		\edef\@p@srwidth{\number\dimen100}
}
\def\@p@@sangle#1{
		\@angletrue
		\edef\@p@sangle{#1} 
}
\def\@p@@ssilent#1{ 
		\@verbosefalse
}
\def\@p@@sprolog#1{\@prologfiletrue\def\@prologfileval{#1}}
\def\@p@@spostlog#1{\@postlogfiletrue\def\@postlogfileval{#1}}
\def\@cs@name#1{\csname #1\endcsname}
\def\@setparms#1=#2,{\@cs@name{@p@@s#1}{#2}}
%
%
\def\ps@init@parms{
		\@bbllxfalse \@bbllyfalse
		\@bburxfalse \@bburyfalse
		\@heightfalse \@widthfalse
		\@rheightfalse \@rwidthfalse
		\def\@p@sbbllx{}\def\@p@sbblly{}
		\def\@p@sbburx{}\def\@p@sbbury{}
		\def\@p@sheight{}\def\@p@swidth{}
		\def\@p@srheight{}\def\@p@srwidth{}
		\def\@p@sangle{0}
		\def\@p@sfile{} \def\@p@sbbfile{}
		\def\@p@scost{10}
		\def\@sc{}
		\@prologfilefalse
		\@postlogfilefalse
		\@clipfalse
		\if@noisy
			\@verbosetrue
		\else
			\@verbosefalse
		\fi
}
%
%
\def\parse@ps@parms#1{
	 	\@psdo\@psfiga:=#1\do
		   {\expandafter\@setparms\@psfiga,}}
%
%
\newif\ifno@bb
\def\bb@missing{
	\if@verbose{
		\ps@typeout{psfig: searching \@p@sbbfile \space  for bounding box}
	}\fi
	\no@bbtrue
	\epsf@getbb{\@p@sbbfile}
        \ifno@bb \else \bb@cull\epsf@llx\epsf@lly\epsf@urx\epsf@ury\fi
}	
\def\bb@cull#1#2#3#4{
	\dimen100=#1 bp\edef\@p@sbbllx{\number\dimen100}
	\dimen100=#2 bp\edef\@p@sbblly{\number\dimen100}
	\dimen100=#3 bp\edef\@p@sbburx{\number\dimen100}
	\dimen100=#4 bp\edef\@p@sbbury{\number\dimen100}
	\no@bbfalse
}
\newdimen\p@intvaluex
\newdimen\p@intvaluey
\def\rotate@#1#2{{\dimen0=#1 sp\dimen1=#2 sp
		  \global\p@intvaluex=\cosine\dimen0
		  \dimen3=\sine\dimen1
		  \global\advance\p@intvaluex by -\dimen3
		  \global\p@intvaluey=\sine\dimen0
		  \dimen3=\cosine\dimen1
		  \global\advance\p@intvaluey by \dimen3
		  }}
\def\compute@bb{
		\no@bbfalse
		\if@bbllx \else \no@bbtrue \fi
		\if@bblly \else \no@bbtrue \fi
		\if@bburx \else \no@bbtrue \fi
		\if@bbury \else \no@bbtrue \fi
		\ifno@bb \bb@missing \fi
		\ifno@bb \ps@typeout{FATAL ERROR: no bb supplied or found}
			\no-bb-error
		\fi
		%
%
		\count203=\@p@sbburx
		\count204=\@p@sbbury
		\advance\count203 by -\@p@sbbllx
		\advance\count204 by -\@p@sbblly
		\edef\ps@bbw{\number\count203}
		\edef\ps@bbh{\number\count204}
		\if@angle 
			\Sine{\@p@sangle}\Cosine{\@p@sangle}
	        	{\dimen100=\maxdimen\xdef\r@p@sbbllx{\number\dimen100}
					    \xdef\r@p@sbblly{\number\dimen100}
			                    \xdef\r@p@sbburx{-\number\dimen100}
					    \xdef\r@p@sbbury{-\number\dimen100}}
%
                        \def\minmaxtest{
			   \ifnum\number\p@intvaluex<\r@p@sbbllx
			      \xdef\r@p@sbbllx{\number\p@intvaluex}\fi
			   \ifnum\number\p@intvaluex>\r@p@sbburx
			      \xdef\r@p@sbburx{\number\p@intvaluex}\fi
			   \ifnum\number\p@intvaluey<\r@p@sbblly
			      \xdef\r@p@sbblly{\number\p@intvaluey}\fi
			   \ifnum\number\p@intvaluey>\r@p@sbbury
			      \xdef\r@p@sbbury{\number\p@intvaluey}\fi
			   }
			\rotate@{\@p@sbbllx}{\@p@sbblly}
			\minmaxtest
			\rotate@{\@p@sbbllx}{\@p@sbbury}
			\minmaxtest
			\rotate@{\@p@sbburx}{\@p@sbblly}
			\minmaxtest
			\rotate@{\@p@sbburx}{\@p@sbbury}
			\minmaxtest
			\edef\@p@sbbllx{\r@p@sbbllx}\edef\@p@sbblly{\r@p@sbblly}
			\edef\@p@sbburx{\r@p@sbburx}\edef\@p@sbbury{\r@p@sbbury}
		\fi
		\count203=\@p@sbburx
		\count204=\@p@sbbury
		\advance\count203 by -\@p@sbbllx
		\advance\count204 by -\@p@sbblly
		\edef\@bbw{\number\count203}
		\edef\@bbh{\number\count204}
}
%
%
\def\in@hundreds#1#2#3{\count240=#2 \count241=#3
		     \count100=\count240	
		     \divide\count100 by \count241
		     \count101=\count100
		     \multiply\count101 by \count241
		     \advance\count240 by -\count101
		     \multiply\count240 by 10
		     \count101=\count240	
		     \divide\count101 by \count241
		     \count102=\count101
		     \multiply\count102 by \count241
		     \advance\count240 by -\count102
		     \multiply\count240 by 10
		     \count102=\count240	
		     \divide\count102 by \count241
		     \count200=#1\count205=0
		     \count201=\count200
			\multiply\count201 by \count100
		 	\advance\count205 by \count201
		     \count201=\count200
			\divide\count201 by 10
			\multiply\count201 by \count101
			\advance\count205 by \count201
		     \count201=\count200
			\divide\count201 by 100
			\multiply\count201 by \count102
			\advance\count205 by \count201
		     \edef\@result{\number\count205}
}
\def\compute@wfromh{
		\in@hundreds{\@p@sheight}{\@bbw}{\@bbh}
		\edef\@p@swidth{\@result}
}
\def\compute@hfromw{
	        \in@hundreds{\@p@swidth}{\@bbh}{\@bbw}
		\edef\@p@sheight{\@result}
}
\def\compute@handw{
		\if@height 
			\if@width
			\else
				\compute@wfromh
			\fi
		\else 
			\if@width
				\compute@hfromw
			\else
				\edef\@p@sheight{\@bbh}
				\edef\@p@swidth{\@bbw}
			\fi
		\fi
}
\def\compute@resv{
		\if@rheight \else \edef\@p@srheight{\@p@sheight} \fi
		\if@rwidth \else \edef\@p@srwidth{\@p@swidth} \fi
}
%
\def\compute@sizes{
	\compute@bb
	\if@scalefirst\if@angle
	\if@width
	   \in@hundreds{\@p@swidth}{\@bbw}{\ps@bbw}
	   \edef\@p@swidth{\@result}
	\fi
	\if@height
	   \in@hundreds{\@p@sheight}{\@bbh}{\ps@bbh}
	   \edef\@p@sheight{\@result}
	\fi
	\fi\fi
	\compute@handw
	\compute@resv}

%
%
\def\psfig#1{\vbox {
	%
	\ps@init@parms
	\parse@ps@parms{#1}
	\compute@sizes
	\ifnum\@p@scost<\@psdraft{
		\special{ps::[begin] 	\@p@swidth \space \@p@sheight \space
				\@p@sbbllx \space \@p@sbblly \space
				\@p@sbburx \space \@p@sbbury \space
				startTexFig \space }
		\if@angle
			\special {ps:: \@p@sangle \space rotate \space} 
		\fi
		\if@clip{
			\if@verbose{
				\ps@typeout{(clip)}
			}\fi
			\special{ps:: doclip \space }
		}\fi
		\if@prologfile
		    \special{ps: plotfile \@prologfileval \space } \fi
		\if@decmpr{
			\if@verbose{
				\ps@typeout{psfig: including \@p@sfile.Z \space }
			}\fi
			\special{ps: plotfile "`zcat \@p@sfile.Z" \space }
		}\else{
			\if@verbose{
				\ps@typeout{psfig: including \@p@sfile \space }
			}\fi
			\special{ps: plotfile \@p@sfile \space }
		}\fi
		\if@postlogfile
		    \special{ps: plotfile \@postlogfileval \space } \fi
		\special{ps::[end] endTexFig \space }
		\vbox to \@p@srheight sp{
			\hbox to \@p@srwidth sp{
				\hss
			}
		\vss
		}
	}\else{
		\if@draftbox{		
			\hbox{\frame{\vbox to \@p@srheight sp{
			\vss
			\hbox to \@p@srwidth sp{ \hss \@p@sfile \hss }
			\vss
			}}}
		}\else{
			\vbox to \@p@srheight sp{
			\vss
			\hbox to \@p@srwidth sp{\hss}
			\vss
			}
		}\fi

	}\fi
}}
\psfigRestoreAt
\let\@=\LaTeXAtSign

%
%
%
\edef\catcodeat{\the\catcode`\@}\catcode`\@=11
\newbox\p@artialpage\newbox\r@emainder
\newbox\b@oxzero\newbox\b@oxone\newbox\b@oxtwo\newbox\b@oxthree
\newdimen\c@oldimen
\newdimen\i@nterskip
\newdimen\p@ageheight
\newdimen\p@agewidth
\newdimen\p@agework
\newdimen\v@dblesize
%
\def\v@oidbox{\hrule height 0pt width 0pt}
\def\b@ottomskip{\vskip 0pt plus 10pt}
\def\dble@safety{0.1}
\def\compute@left{\p@agework=\p@ageheight
 \ifvoid\footins\relax\else
   \advance\p@agework by -\ht\footins
   \advance\p@agework by -\skip\footins
 \fi 
 \ifvoid\topins\relax\else
   \advance\p@agework by -\ht\topins
   \advance\p@agework by -\skip\topins
 \fi}%
%
\def\out@sofar{{\def\pagecontents{\ifvoid\topins\else\unvbox\topins\fi
   \p@agesofar
   \ifvoid\footins\else 
     \vskip\skip\footins
     \footnoterule
     \unvbox\footins
   \fi}\hsize=\p@agewidth\vsize=\p@ageheight
   \givenoutput}}
%
\def\out@partial{{\def\pagecontents{\ifvoid\topins\else\unvbox\topins\fi
   \unvbox\p@artialpage
   \ifvoid\footins\else 
     \vskip\skip\footins
     \footnoterule
     \unvbox\footins
   \fi}\hsize=\p@agewidth\vsize=\p@ageheight
   \givenoutput}}
\def\safe@eject{\vsize=\p@ageheight\pagegoal=4\p@ageheight\eject
  \removelastskip}
%
\def\begindoublecolumns{\p@ageheight=\vsize\p@agewidth=\hsize
\edef\givenoutput{\the\output}%
\edef\restoretopskip{\noexpand
   \topskip=\the\topskip\noexpand\splittopskip=\the\splittopskip}%
\topskip=\baselineskip\splittopskip=\topskip
\edef\restorebaselineskip{\noexpand\baselineskip=\the\baselineskip
   \noexpand\normallineskip=\the\normallineskip}%
%
 \compute@left
 \advance\p@agework by -2\baselineskip
 \ifdim\pagetotal > \p@agework
   \message{Begindoublecolumns: Pagetotal=\the\pagetotal\ >
   P@agework=\the\p@agework}\relax
   {\splittopskip=\topskip\splitmaxdepth=\maxdepth
   \output{\global\setbox\b@oxone=\vbox{\unvbox255}}\safe@eject
   \global\setbox\p@artialpage=\vsplit\b@oxone to \p@agework}\out@partial
   \unvbox\b@oxone
 \fi
 \begingroup\multiply\hyphenpenalty by 2\relax
   \widowpenalty 60\relax
   \clubpenalty 30\relax
   \brokenpenalty 50\relax
   \linepenalty 50\relax
   \output={\global\setbox\p@artialpage=\vbox{\unvbox255}}\relax
   \baselineskip=0pt \normallineskip=0pt    
   \boxmaxdepth=0pt\maxdepth=0pt\relax
   \topskip=0pt\splittopskip=0pt\safe@eject
   \restorebaselineskip
   \compute@left \advance\p@agework by \p@ageheight 
   \v@oidbox
   \global\c@oldimen=2\ht\p@artialpage
   \global\advance\c@oldimen by -\dble@safety\ht\p@artialpage
   \global\advance\c@oldimen by -\dble@safety\ht\p@artialpage
   \global\advance\c@oldimen by \dble@safety\p@agework
   \v@dblesize=\p@agework
   \advance\v@dblesize by -\c@oldimen
   \ifdim\v@dblesize < \baselineskip
     \message{ P@agework=\the\p@agework, C@oldimen=\the\c@oldimen,
       V@dblesize=\the\v@dblesize}
   \fi  
   \output={\doublecolumnout}\hsize=0.48\p@agewidth\vsize=2\p@ageheight
 }
%
\def\enddoublecolumns{\removelastskip
 \compute@left\advance\p@agework by \p@ageheight
 \advance\v@dblesize by -2\baselineskip
 \ifdim\pagetotal > \v@dblesize
   \output{\global\setbox\r@emainder=\vbox{\unvbox255}}\safe@eject
   \splittopskip=\topskip\boxmaxdepth=\maxdepth
   \global\setbox255=\vsplit\r@emainder to \v@dblesize
   \doublecolumnout
   \unvbox\r@emainder
 \fi   
 \output={\balancecolumns}\safe@eject
 \endgroup 
 \vsize=\p@ageheight
 \v@oidbox
 \pagegoal=2\p@ageheight\penalty 10000\p@agesofar
 \vsize=\p@ageheight
 \v@oidbox\pagegoal=\p@ageheight\par\penalty -2000}%
\def\doublecolumnout{\splittopskip=\topskip\splitmaxdepth=\maxdepth
 \vsize=\p@ageheight
 \compute@left
 \advance\p@agework by -\ht\p@artialpage
 \global\c@oldimen=2\p@agework
 \global\setbox\b@oxzero=\vsplit255 to \c@oldimen
 \balanceb@oxzero
 \out@sofar
 \vsize=2\p@ageheight\pagegoal=\vsize
 \unvbox255\restoretopskip
 \vsize=2\p@ageheight\pagegoal=\vsize
 \penalty\outputpenalty
}%
\def\p@agesofar{\unvbox\p@artialpage\penalty 10000\relax
 \wd\b@oxzero=\hsize \wd\b@oxtwo=\hsize
 \hbox to \p@agewidth{\vtop{\unvbox\b@oxzero}\hfil\vtop{\unvbox\b@oxtwo}}}
%
\def\balanceb@oxzero{\global\c@oldimen=\ht\b@oxzero
 \global\advance\c@oldimen by \topskip
 \global\advance\c@oldimen by -\baselineskip
 \global\divide\c@oldimen by 2\relax
 \splittopskip=\topskip\splitmaxdepth=\maxdepth
{\vbadness=10000\loop
\global\setbox\b@oxthree=\copy\b@oxzero
\global\setbox\b@oxtwo=\vsplit\b@oxthree to \c@oldimen
\global\setbox\b@oxone=\vbox{\unvbox\b@oxtwo}%
\p@agework=\ht\b@oxthree
\advance\p@agework by -\ht\b@oxone
\ifdim\p@agework > \baselineskip
 \global\advance\c@oldimen by 0.2\p@agework
\repeat}\relax
{\vbadness=10000\loop
\global\setbox\b@oxthree=\copy\b@oxzero
\global\setbox\b@oxtwo=\vsplit\b@oxthree to \c@oldimen
\global\setbox\b@oxone=\vbox{\unvbox\b@oxtwo}%
\p@agework=\ht\b@oxthree
\advance\p@agework by -\ht\b@oxone
\ifdim\p@agework < -\baselineskip
 \global\advance\c@oldimen by 0.2\p@agework
\repeat}\relax
\p@agework=\ht\b@oxthree
\advance\p@agework by -\ht\b@oxone
\ifdim\p@agework > \baselineskip
  \message{SPLIT: p.\the\count0, P@gework: \the\p@agework}\relax
  \i@nterskip=\baselineskip
  \global\setbox\b@oxtwo=\hbox{\rm W}\advance\i@nterskip by -\ht\b@oxtwo
  \global\setbox\b@oxtwo=\vsplit\b@oxthree to 0.5\p@agework
\else
  \i@nterskip=\topskip
\fi
\global\setbox\b@oxzero=\vbox to \c@oldimen{\unvbox\b@oxone
 \vskip\i@nterskip \unvbox\b@oxtwo\b@ottomskip}%
\global\setbox\b@oxtwo=\vbox to \c@oldimen{\unvbox\b@oxthree\b@ottomskip}%
}%
\def\balancecolumns{\global\setbox\b@oxzero=\vbox{\unvbox255}\balanceb@oxzero
}%
%
\def\leftrule{\removelastskip\par\kern -5pt\hbox{\vrule width 0.49\hsize height 0.4pt
\vrule height 6pt}\medskip}
\def\rightrule{\medskip\rightline{\vrule depth 5.6pt height 0.4pt width 0.4pt
\vrule width 0.49\hsize height 0.4pt depth 0pt}\kern -5pt\par}
\catcode`\@=\catcodeat
%

\hsize=7.0in

\def\nWm2sr{~{\rm nW~m}^{-2}~{\rm sr}^{-1}}
\def\etal{{\rm~et~al.~}}
\def\altaffilmark#1{$^{#1}$}

\journal{}
\vskip-30pt
\title{THE COBE DIFFUSE INFRARED BACKGROUND EXPERIMENT} \vskip-6pt
\titlem{SEARCH FOR THE COSMIC INFRARED BACKGROUND:}
\titlem{I. LIMITS AND DETECTIONS}
\author{M.G.~Hauser\altaffilmark{1},
        R.G.~Arendt\altaffilmark{2},
        T.~Kelsall\altaffilmark{3},
        E.~Dwek\altaffilmark{3},
        N.~Odegard\altaffilmark{2},
        J.L.~Weiland\altaffilmark{2},
        H.T.~Freudenreich\altaffilmark{2},}
\author{W.T.~Reach\altaffilmark{4},
        R.F.~Silverberg\altaffilmark{3},
        S.H.~Moseley\altaffilmark{3},
        Y.C.~Pei\altaffilmark{1},
        P.~Lubin\altaffilmark{5},
        J.C.~Mather\altaffilmark{3},
        R.A.~Shafer\altaffilmark{3},}
\author{G.F.~Smoot\altaffilmark{6},
        R.~Weiss\altaffilmark{7},
        D.T.~Wilkinson\altaffilmark{8}, and
        E.L.~Wright\altaffilmark{9}}
\author{{\it Received 1998 January 6; accepted 1998 June 3}}
\abstract
The Diffuse Infrared Background Experiment (DIRBE) on the Cosmic
Background Explorer ({\it COBE}) spacecraft was designed primarily
to conduct a systematic search for an isotropic cosmic infrared
background (CIB) in ten photometric bands from 1.25 to 240~$\mu$m.
The results of that search are presented here.  Conservative limits
on the CIB are obtained from the minimum observed brightness in
all-sky maps at each wavelength, with the faintest limits in the DIRBE
spectral range being at 3.5~$\mu$m $(\nu I_\nu < 64 \nWm2sr$, 95\% CL)
and at 240~$\mu$m $(\nu I_\nu < 28 \nWm2sr$, 95\% CL).  The bright
foregrounds from interplanetary dust scattering and emission, stars,
and interstellar dust emission are the principal impediments to the
DIRBE measurements of the CIB.  These foregrounds have been modeled
and removed from the sky maps. Assessment of the random and systematic
uncertainties in the residuals and tests for isotropy show that only
the 140 and 240 $\mu$m data provide candidate detections of the CIB. 
The residuals and their uncertainties provide CIB upper limits more
restrictive than the dark sky limits at wavelengths from 1.25 to 100
$\mu$m.  No plausible solar system or Galactic source of the observed
140 and 240 $\mu$m residuals can be identified, leading to the
conclusion that the CIB has been detected at levels of $\nu I_\nu=
25\pm7$ and $14\pm3 \nWm2sr$ at 140 and 240 $\mu$m respectively.
The integrated energy from 140 to 240 $\mu$m, 10.3 $\nWm2sr$, is about
twice the integrated optical light from the galaxies in the Hubble
Deep Field, suggesting that star formation might have been heavily
enshrouded by dust at high redshift.  The detections and upper limits
reported here provide new constraints on models of the history of
energy-releasing processes and dust production since the decoupling of
the cosmic microwave background from matter.
\subject
cosmology: observations --- diffuse radiation --- infrared: general

\maintext
\section{1. INTRODUCTION}

The search for the cosmic infrared background (CIB) radiation is a
relatively new field of observational cosmology.  The term ``CIB''
itself has been used with various meanings in the literature; we
define it here to mean all diffuse infrared radiation arising external
to the Milky Way Galaxy.  Measurement of this distinct radiative
background, expected to arise from the cumulative\break

{\baselineskip=11pt\eightpoint
\par$^1$ Space Telescope Science Institute, 3700 San Martin Drive,
Baltimore, MD 21218
\par$^2$ Raytheon STX, Code 685, NASA Goddard Space Flight Center,
Greenbelt, MD 20771
\par$^3$ Code 685, NASA Goddard Space Flight Center, Greenbelt, MD
20771
\par$^4$ California Institute of Technology, IPAC/JPL, MS 100-22,
Pasadena, CA 91125
\baselineskip=12pt}

\noindent
emissions of pregalactic, protogalactic, and evolved galactic systems,
would provide new insight into the cosmic ``dark ages'' following the
decoupling of matter from the cosmic microwave background (CMB)
radiation (Partridge \& Peebles 1967; Harwit 1970; Bond, Carr, \&
Hogan 1986, 1991; Franceschini et al. 1991, 1994; Fall, Charlot, \&
Pei 1996).

The search for the CIB is impeded by two fundamental challenges: there
is no unique spectral signature of\break

{\baselineskip=11pt\eightpoint
\par$^5$ UCSB, Physics Department, Santa Barbara, CA 93106
\par$^6$ Lawrence Berkeley Laboratory, Space Sciences
Laboratory, Department of Physics, UC Berkeley, CA 94720
\par$^7$ Massachusetts Institute of Technology, Room 20F-001,
Department of Physics, Cambridge, MA 02139
\par$^8$ Princeton University, Department of Physics, Jadwin Hall,
Box 708, Princeton, NJ 08544
\par$^9$ UCLA, Astronomy Department, Los Angeles, CA 90024
\baselineskip=12pt}

\noindent
such a background, and there are many local contributors to the
infrared sky brightness at all wavelengths, several of them quite
bright.  The lack of a distinct spectral signature arises in part
because so many different sources of primordial luminosity are
possible (e.g., Bond, Carr, \& Hogan 1986), in part because the
radiant characteristics of evolving galaxies are imperfectly known,
and in part because the primary emissions at any epoch are then
shifted into the infrared by the cosmic red-shift and possibly by dust
absorption and re-emission.  Hence, the present spectrum depends in a
complex way on the characteristics of the luminosity sources, on their
cosmic history, and on the dust formation history of the Universe.

Setting aside the difficult possibility of recognizing the CIB by its
angular fluctuation spectrum (Bond, Carr, \& Hogan 1991; Kashlinsky et
al. 1996a, b), the only identifying CIB characteristic for which one
can search is an isotropic signal.  Possible evidence for an isotropic
infrared background, or at least limits on emission in excess of local
foregrounds, has been reported on the basis of very limited data from
rocket experiments (Matsumoto et al. 1988; Matsumoto 1990; Noda et
al. 1992; Kawada et al. 1994).  Puget et al. (1996) have used data
from the {\it COBE} Far Infrared Absolute Spectrophotometer (FIRAS) to
conclude that there is tentative evidence for a CIB at submillimeter
wavelengths.  Indirect upper limits, and even possible lower limits,
on the extragalactic infrared background have been inferred from the
apparent attenuation of TeV $\gamma$-rays in propagation from distant
sources (de Jager, Stecker, \& Salamon 1994; Dwek \& Slavin 1994;
Biller et al. 1995; Stecker 1996; Stecker \& de Jager 1997).  However,
the detection of TeV $\gamma$-rays from Markarian 421 recently
reported by Krennrich et al. (1997) casts some doubt on the infrared
background inferred in this manner.  The integrated energy density of
the CIB in units of the critical density might, on the basis of
pre-{\it COBE} observations (Ressell \& Turner 1990), exceed that of
the CMB, $\Omega_{\rm CMB}=1\times10^{-4}h_{50 }^{-2}$, and
preliminary DIRBE results (Hauser 1995, 1996a, b) only set limits on
the integrated CIB which are comparable to the energy density in the
CMB.

Direct detection of the CIB requires a number of steps.  One must
solve the formidable observational problem of making absolute
brightness measurements in the infrared.  One must then discriminate
and remove the strong signals from foregrounds arising from one's
instrument or observing environment, the terrestrial atmosphere, the
solar system, and the Galaxy.  Particular attention must be given to
possible isotropic contributions from any of these foreground sources.

This paper summarizes the results of the DIRBE investigation, in which
a direct measurement of the CIB has been made by measuring the
absolute sky brightness at ten infrared wavelengths and searching for
isotropic radiation arising outside of the solar system and Galaxy.
We report upper limits on the CIB from 1.25 to 100~ $\mu$m, and
detection of the CIB at 140 and 240 $\mu$m.  Section 2 briefly
describes the DIRBE instrument and the character of its data.  Section
2 also summarizes the procedures used to model foreground radiations,
and for estimating the random and systematic uncertainties in the
measurements and the models.  Because the foreground models are
critical to our conclusions, they are also described more extensively
in separate papers.  Details of the interplanetary dust (IPD) model
used to discriminate the sky brightness contributed by dust in the
solar system are provided by Kelsall et al. (1998, hereafter Paper
II).  Arendt et al. (1998, hereafter Paper III) describe the Galactic
foreground discrimination procedures and summarize systematic errors
in the foreground determination process.  Section 3 of this paper
summarizes the observational results, presented in compact form in
Table 2.  Dwek et al. (1998, hereafter Paper IV) show in detail that
the isotropic residuals detected at 140 and 240 $\mu$m are not likely
to arise from unmodeled solar system or Galactic sources.  Section 4
of this paper summarizes that analysis, provides a comparison of the
DIRBE results with other diffuse brightness and integrated discrete
source measurements, presents limits on the integrated energy in the
cosmic infrared background implied by the DIRBE measurements, and
briefly discusses the implications of these results for models of
cosmic evolution.  A more extensive discussion of the implications is
provided in Paper IV.  Independent confirmation of the DIRBE
observational results and extension of the CIB detection to longer
wavelengths is provided by Fixsen et al. (1998), as discussed in
\S~4.2.1.  The remainder of this Section provides an overview of the
rather extensive arguments presented in this paper as a guide to the
reader.

From absolute brightness maps of the entire sky over ten months of
observation, the faintest measured value at each wavelength is
determined (\S~3.1 and Table 2).  These ``dark sky'' values are either
direct measurements of the CIB (if we were fortuitously located in the
Universe), or yield conservative upper limits on it.  Since the
measured infrared sky brightness is not isotropic at any wavelength in
the DIRBE range, it cannot be concluded that these dark sky values are
direct detections of the CIB.  As expected, the dark sky values are
least near 3.5 $\mu$m, in the relative minimum between scattering of
sunlight by interplanetary dust and re-emission of absorbed sunlight
by the same dust, and at the longest DIRBE wavelength, 240 $\mu$m,
where emission from interstellar dust is decreasing from its peak at
shorter wavelengths and the cosmic microwave background has not yet
become significant.

To proceed further, the contributions from the solar system and Galaxy
to the DIRBE maps are determined.  The contribution of interplanetary
dust is recognizable because motion of the Earth in its orbit through
this cloud causes annual variation of the sky brightness in all
directions.  An empirical, parametric model of the IPD cloud (\S~2.3
and Paper II) is used to extract the IPD contribution.  Though this
model is not unique, Paper II demonstrates that the implications for
the CIB are reasonably robust, that is, rather insensitive to
variations in the model.

\hskip-0.94pt
The Galactic contribution from discrete sources bright enough to be
detected individually is simply deleted from further analysis by
blanking a small surrounding region in the maps.  The integrated
contributions of faint discrete Galactic sources are calculated at
each wavelength from 1.25 to 25 $\mu$m from a detailed statistical
model of Galactic sources and their spatial distribution (\S~2.3 and
Paper III).  The contribution from the diffuse interstellar medium
(ISM) at each wavelength is obtained by scaling a template map of ISM
emission to that wavelength.  At all wavelengths except 100 $\mu$m,
the template is the residual 100 $\mu$m map after removal of the IPD
contribution, a map where ISM emission is prominent.  To remove the
ISM contribution without removing some fraction of the CIB at other
wavelengths, the 100 $\mu$m extragalactic light is first estimated by
extrapolating the ${\rm H~I}-100~\mu$m correlation to zero H~I column
density for two fields, the Lockman hole (Lockman, Jahoda, \& McCammon
1986) and north ecliptic pole, where there is known to be little other
interstellar gas (molecular or ionized) in the line of sight (\S~3.4).
This estimate is subtracted from the 100 $\mu$m map before scaling it
to other wavelengths.  The ISM template at 100 $\mu$m was chosen to be
the map of H~I emission, scaled by the slope of the H~I-to-100 $\mu$m
correlation (\S~2.3 and Paper III).

Clearly, drawing the proper conclusions from the DIRBE measurements
and foreground models is critically dependent upon assessment of the
uncertainties in both the measurements and the models.  These
uncertainties are discussed at length in Papers II and III, and are
summarized here in \S~2.4 and Table 2.

Because the foreground emissions are so bright, the definitive search
for evidence of the CIB is carried out on the residual maps after
removal of the solar system and Galactic foregrounds in a restricted
region of the sky at high galactic and ecliptic latitudes (designated
the ``high quality'' region B, HQB, discussed in \S~3.3 and defined in
Table~3).  The HQB region is the largest area in which the residual
maps do not clearly contain artifacts from the foreground removal, and
covers about 2\% of the sky.  It includes regions in both the northern
and southern hemispheres and allows isotropy testing on 8,140 map
pixels over angular scales up to 43 degrees within each hemisphere and
from 137 to 180 degrees between hemispheres.  In this region, the mean
residuals are determined and their uncertainties are estimated.  More
precise estimates of the mean residuals at 100, 140 and 240 $\mu$m are
obtained from a weighted average of values determined in the HQB
region and in well-studied faint regions toward the Lockman hole and
the north ecliptic pole.  The residuals are tested for significance by
requiring that they exceed three times the estimated uncertainty
including both random and systematic effects.

The final step toward recognition of the CIB is to test for isotropy
of the residuals.  Though a number of approaches are discussed
(\S~3.5), the conclusions are finally based upon the absence of
significant spatial correlations of the residuals with any of the
foreground models or with galactic or ecliptic latitude and the
absence of significant structure in the 2-point correlation function
in the HQB region.

Only at 140 and 240 $\mu$m do the results meet our two necessary
criteria for CIB detection: significant residual in excess of
3$\sigma$ and isotropy in the HQB region (\S~3.6).  These isotropic
residuals are unlikely to arise from unmodeled solar system or
Galactic sources (\S~4.1 and Paper IV), leading to the conclusion that
the CIB has been detected at 140 and 240 $\mu$m.  At each wavelength
shorter than 100 $\mu$m, an upper limit to the CIB is set at 2$\sigma$
above the mean HQB residual, which in all cases is a more restrictive
limit than the dark sky limit.  At 100 $\mu$m, the most restrictive
limit is found from the weighted average of the residuals in the HQB
region, the Lockman hole and the north ecliptic pole.  The last line
of Table 2 shows the final CIB limits and detected values.

\section{2. DIRBE, DATA, AND PROCEDURES}

This section provides a brief review of the important features of the
DIRBE instrument, the data it provides, and our reduction of the data
with the goal of extracting the CIB.  These topics are more thoroughly
described in the {\it COBE}/DIRBE Explanatory Supplement (1997), and
Papers II and III.

\subsection{2.1. DIRBE Instrument Description}

The {\it COBE} Diffuse Infrared Background Experiment was the first
satellite instrument designed specifically to carry out a systematic
search for the CIB in the $1.25-240$~$\mu$m range.  A detailed
description of the {\it COBE} mission has been given by Boggess et
al. (1992), and the DIRBE instrument has been described by Silverberg
et al. (1993).  The DIRBE observational approach was to obtain
absolute brightness maps of the full sky in 10 broad photometric bands
at 1.25, 2.2, 3.5, 4.9, 12, 25, 60, 100, 140, and 240 $\mu$m.  Table~1
summarizes the instrumental parameters, including the effective band
width, beam solid angle, detector type, and filter construction.
Though linear polarization was also measured at 1.25, 2.2, and 3.5
$\mu$m, the polarization information has not been used in this
analysis.

DIRBE characteristics of particular relevance to the CIB search
include:

(a) Highly redundant sky coverage over a range of elongation angles.
    Because the diffuse infrared brightness of the entire sky varies
    as a result of our motion within the IPD cloud (and possible
    variations of the cloud itself), the DIRBE was designed to scan
    half the sky every day, providing detailed ``light curves'' with
    hundreds of samples over the mission for every pixel.  This
    sampling provides a strong means of discriminating solar system
    emission.  The scanning was produced by offsetting the DIRBE
    line-of-sight by $30^\circ$ from the {\it COBE} spin axis, which
    was normally fixed at a solar elongation angle of $94^\circ$,
    providing sampling at elongation angles ranging from $64^\circ$ to
    $124^\circ$.  Such sampling also modulates the signal from any
    nearby spherically symmetric Sun-or Earth-centered IPD component,
    which would otherwise appear as a constant (i.e., ``isotropic'')
    signal.

(b) Sensitivity. Table 2 lists one-sigma instrumental sensitivities,
    $\sigma(\nu I_\nu)$, for each $0\fdg7\times0\fdg7$ field of view
    over the complete 10 months of cryogenic operation.  These single
    field-of-view values are generally below the actual sky
    brightness, and below many of the predictions for the CIB.
    Averaging over substantial sky areas, once foregrounds are
    removed, increases the sensitivity for an isotropic signal.

\def\s{&}\def\nl{\cr}\def\tablenotemark#1{$^{#1}$}
\footnote{}{\eightpoint
 \tabbot{1}{DIRBE I{\sc NSTRUMENT} C{\sc HARACTERISTICS}}
 {
  \hfil#\hfil&\hfil#\hfil&\hfil#\hfil&\hfil#\hfil&
  \hfil#\hfil&\hfil#\hfil&\hfil#\hfil&\hfil#\hfil\cr
  \tabhead{
   {Band}&{$\lambda$\tablenotemark{a}}&{$\Delta\nu_{e}$\tablenotemark{b}}&
   {Beam Solid Angle}&{Detector Type}&{Filter Construction\tablenotemark{c}}&
   {Absolute Calibration}\nl
   {}&{($\micron$)}&{(Hz)}&{($10^{-4}$ sr)}&{}&{}&{Reference Source}\nl}
  \tabbody{
1 \s 1.25\tablenotemark{d} \s $5.95\times10^{13}$ \s 1.198 \s 
          InSb\tablenotemark{e} \s Coated Glass \s Sirius \nl
2 \s  2.2\tablenotemark{d} \s $2.24\times10^{13}$ \s 1.420 \s
          InSb\tablenotemark{e} \s Coated Glass \s Sirius \nl
3 \s  3.5\tablenotemark{d} \s $2.20\times10^{13}$ \s 1.285 \s
          InSb\tablenotemark{e} \s Coated Germanium  \s Sirius \nl
4 \s  4.9       \s $8.19\times10^{12}$ \s 1.463 \s InSb$^{e}$ \s
          MLIF/Germanium  \s Sirius \nl
5 \s   12       \s $1.33\times10^{13}$ \s 1.427 \s Si:Ga BIB \s
          MLIF/Germanium/ZnSe \s Sirius \nl
6 \s   25       \s $4.13\times10^{12}$ \s 1.456 \s Si:Ga BIB \s
          MLIF/Silicon \s NGC 7027 \nl
7 \s   60       \s $2.32\times10^{12}$ \s 1.512 \s Ge:Ga \s
          MLIF/Sapphire/KRS5/Crystal Quartz \s Uranus \nl
8 \s  100       \s $9.74\times10^{11}$ \s 1.425 \s Ge:Ga \s
          MLIF/KCl/CaF$_2$/Sapphire \s Uranus \nl
9 \s  140       \s $6.05\times10^{11}$ \s 1.385 \s Si/diamond bolometer \s
          Sapphire/Mesh Grids/BaF$_2$/KBr \s Jupiter \nl
10\s  240       \s $4.95\times10^{11}$ \s 1.323 \s Si/diamond bolometer \s 
          Sapphire/Grids/BaF$_2$/CsI/AgCl \s Jupiter \nl}
 }
 \tabnote{$^a$ Nominal wavelength of DIRBE band.}
 \vskip18pt
 \tabnote{$^b$ Effective bandwidth assuming source spectrum $\nu I_{\nu}$
= constant.}
 \vskip18pt
 \tabnote{$^c$ MLIF = multi--layer interference filter.}
 \vskip18pt
 \tabnote{$^d$ Linear polarization and total intensity measured.}
 \vskip18pt
 \tabnote{$^e$ Anti--reflection coated for the band center wavelength.}
}

(c) Stray light rejection.  The DIRBE optical configuration (Magner
    1987) was carefully designed for strong rejection of stray light
    from the Sun, Earth limb, Moon or other off-axis celestial
    radiation, as well as radiation from other parts of the {\it COBE}
    payload (Evans 1983).  Extrapolations of the off-axis response to
    the Moon indicate that stray light contamination for a single
    field-of-view in faint regions of the sky does not exceed
    $1\nWm2sr$ at any wavelength ({\it COBE}/DIRBE Explanatory
    Supplement 1997).

(d) Instrumental offsets. The instrument, which was maintained at a
    temperature below 2 K within the {\it COBE} superfluid helium
    dewar, measured absolute brightness by chopping between the sky
    signal and a zero-flux internal reference at 32 Hz.  Instrumental
    offsets were measured about five times per orbit by closing a cold
    shutter located at the prime focus.  A radiative offset signal in
    the long wavelength detectors arising from JFETs (operating at
    about 70 K) used to amplify the detector signals was identified
    and measured in this fashion and removed from the DIRBE data.
    Because the offset signal was stable over the course of the
    mission, it would appear as an isotropic signal if left
    uncorrected.  To establish the origin of the radiative offset
    signal, and determine whether its value was the same whether the
    instrument shutter was closed (when the offset was monitored) or
    open (when the sky brightness plus offset was measured), special
    tests were conducted during two one-week periods of the mission.
    In these tests, power to individual JFETs was turned off
    sequentially while measuring the offset (shutter closed) and sky
    brightness (shutter open) with all remaining operating detectors.
    The sky brightness measurements at each wavelength with JFETs off
    and on at other wavelengths were carefully compared.  The offsets
    measured in this fashion were consistent with those measured by
    closing the shutter in normal operations, demonstrating that
    changing the position of the shutter did not significantly modify
    the offset.  The final uncertainties in the offset corrections,
    shown as $S({\rm offset})$ in Table 2, are dominated by the
    uncertainties in the results of these special tests, due to the
    limited amount of time devoted to them.  The uncertainties are
    quite negligible at wavelengths less than 140 $\mu$m.  The
    accuracy of the DIRBE measurement zero point at 140 and 240
    $\mu$m, where the offset uncertainty exceeds $1 \nWm2sr$, has been
    independently confirmed by comparison with {\it COBE}/FIRAS data,
    as discussed below.

(e) Gain stability. Short-term stability and linearity of the
    instrument response were monitored using internal radiative
    reference sources which were used to stimulate all detectors each
    time the shutter was closed.  The highly redundant sky sampling
    allowed the use of stable celestial sources to provide precise
    photometric closure over the sky and reproducible photometry to
    $\sim$ 1\% or better for the duration of the mission.

(f) Absolute gain calibration. Calibration of the DIRBE photometric
    scale was obtained from observations of a few isolated bright
    celestial sources ({\it COBE}/ DIRBE Explanatory Supplement 1997).
    Table 1 lists the DIRBE gain reference sources, and Table 2 lists
    the uncertainties in the absolute gain, $S({\rm gain})$, for each
    DIRBE spectral band.

An independent check of the DIRBE offset and absolute gain
calibrations at 100, 140 and 240 $\mu$m has been performed by Fixsen
et al. (1997) using data taken concurrently by the FIRAS instrument on
board {\it COBE}.  The FIRAS calibration is intrinsically more
accurate than that of the DIRBE, but the FIRAS sensitivity drops
rapidly at wavelengths shorter than $200~\mu$m, effectively only
partially covering the DIRBE 100 $\mu$m bandpass.  In general, the two
independent calibrations are consistent within the estimated DIRBE
uncertainties.  Quantitatively, Fixsen et al.  evaluated the gain and
offset corrections needed to bring the two sets of measurements into
agreement.  Taking account of the absolute FIRAS calibration
uncertainty and the uncertainty arising from the comparison process
itself (due in part to the need to integrate the FIRAS data over the
broad DIRBE spectral response in each band and to integrate the DIRBE
data over the large FIRAS beam shape to obtain comparable maps),
Fixsen et al. (1997) found statistically significant, but small,
corrections ($3\sigma$ or greater) to the DIRBE calibration only at
240 $\mu$m.  All results in this paper are based upon the DIRBE
calibration and its uncertainties. The small effect of adopting the
FIRAS calibration at 140 and 240 $\mu$m, which has no qualitative
effect on the conclusions presented here, is discussed in \S~4.2.1.

\subsection{2.2. The DIRBE Data}

The calibrated DIRBE photometric observations are made into maps of
the sky by binning each sample into a pixel on the {\it COBE} sky cube
projection in geocentric ecliptic coordinates ({\it COBE}/DIRBE
Explanatory Supplement 1997).  The projection is nearly equal-area and
avoids geometrical distortions at the poles. Pixels are roughly
20$\arcmin$ on a side.  Forty-one Weekly Maps have been produced by
forming a robust average of all observations of each pixel taken
during a week.  About one-half of the sky is covered each week;
complete sky coverage is achieved within four months.  Data used in
this analysis originate from the Weekly Sky Maps produced by the 1996
Pass 3b DIRBE pipeline software, as documented in the {\it COBE}/DIRBE
Explanatory Supplement (1997).

All analysis is performed on maps in the original sky-cube coordinate
system.  For illustrational purposes, the maps shown in Figure 1 of
this paper are reprojected into an azimuthal equal-area projection.
The DIRBE surface brightness maps are stored as $I_{\nu}$ in units of
MJy~sr$^{-1}$.  Many of the results in this paper are presented as
$\nu I_{\nu}$, where $\nu I_{\nu}({\rm nW~m}^{-2}~{\rm sr}^{-1})=
(3000~\mu{\rm m}/\lambda)I_{\nu}({\rm MJy~sr}^{-1})$.

\subsection{2.3. Foreground Removal Procedures}

Conservative upper limits on the CIB are easily determined from the
minimum sky signal observed at each wavelength; these results are
quoted in \S~3.1.  In order to derive more interesting limits or
detections, one must address the problem of discriminating the various
contributions to the measured sky brightness.  The procedures used to
discriminate and remove foreground emissions from the solar system and
Galaxy are carefully based on distinguishing observational
characteristics of these sources.  Isotropy of the residuals was not
assumed or imposed, but was rigorously tested (\S~3.5).

The approach adopted here is to derive, for each DIRBE wavelength,
$\lambda$, an all-sky map of the residual intensity $I_{res}$
remaining after the removal of solar-system and Galactic foregrounds
from the observed sky brightness $I_{obs}$:
$$
I_{res}(l,b,\lambda)=I_{obs}(l,b,\lambda,t)-Z(l,b,\lambda,t)-
G(l,b,\lambda), \eqno(1)
$$
where $l$ and $b$ are galactic longitude and latitude, $t$ is\break

{\topinsert\null\vskip0pt
  \psfig{file=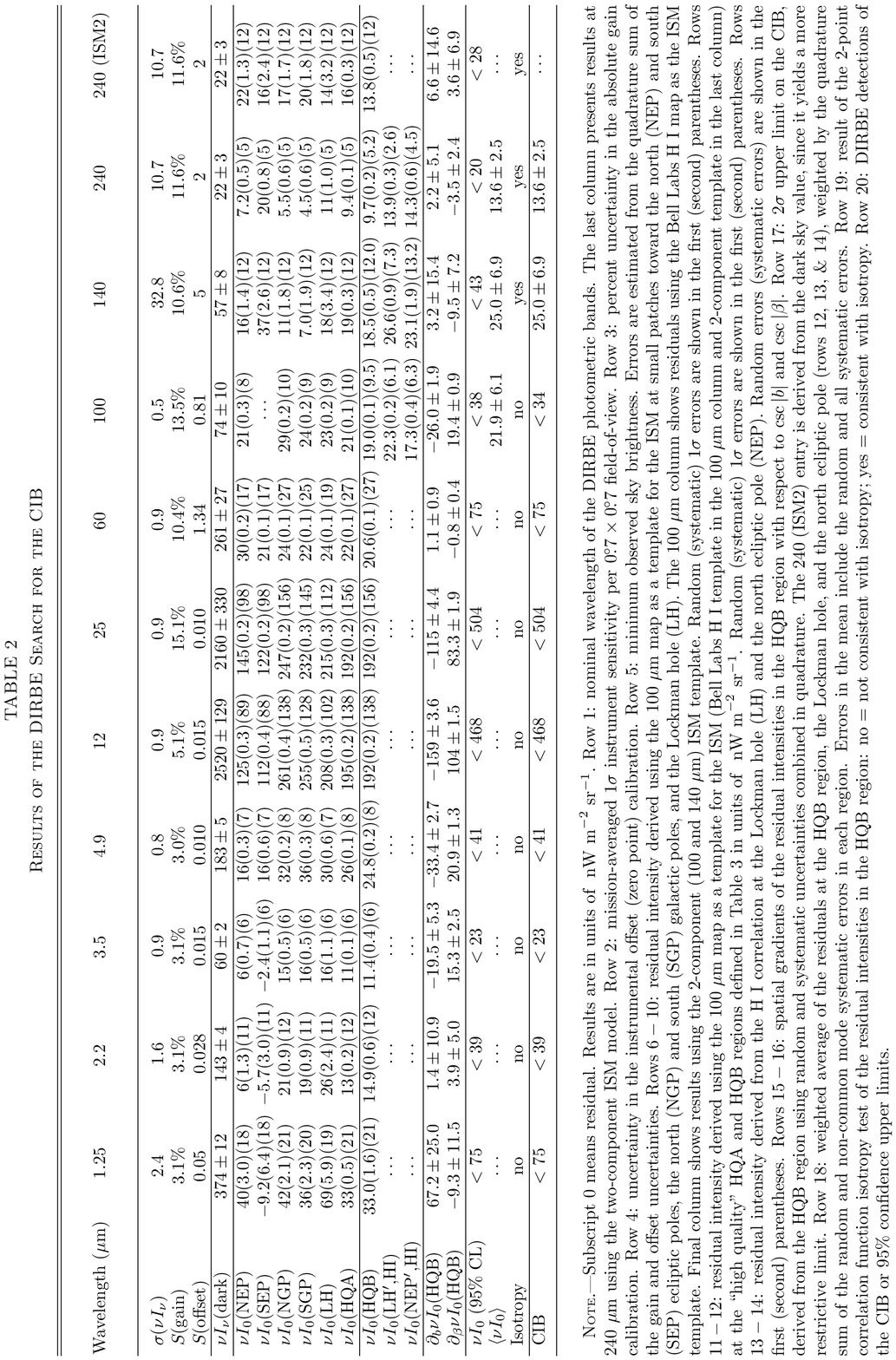,width=\h@size,angle=180,silent=yes}
  \edef\h@old{\the\hsize}\hsize=\h@size
  \hsize=\h@old\endinsert\vfuzz=0.5pt}

\noindent
time,
$Z(l,b,\lambda,t)$ is the contribution from the interplanetary dust
cloud, and $G(l,b,\lambda)$ is the contribution from both stellar and
interstellar dust components within the Galaxy.  Both $Z$ and $G$ are
derived from models.  The choice of models is motivated by the primary
goal of ensuring that no part of the CIB is inadvertently included in
the interplanetary dust cloud or Galactic emission components.
Figure~1 presents maps of $I_{res}$ as derived from the foreground
removal process.

The DIRBE IPD model (Paper II) is a semi-physical, parametric model of
the sky brightness similar, but not identical, to that used to create
the {\it IRAS} Sky Survey Atlas (Wheelock et al. 1994).  The model
represents the sky brightness as the integral along the line-of-sight
of the product of an emissivity function and a three-dimensional dust
density distribution function.  The emissivity function includes both
thermal emission and scattering.  The thermal emission at each
location assumes a single dust temperature for all cloud components.
The temperature is a function only of distance from the Sun and varies
inversely as a power law with distance.  The density distribution
includes a smooth cloud, three pairs of asteroidal dust bands, and a
circumsolar dust ring.  The model is intrinsically static, except that
structure within the circumsolar ring near 1 AU is assumed to co-orbit
the Sun with the Earth.  The apparent seasonal brightness variation
arises from the motion of the Earth on an eccentric orbit within the
cloud, which is not required to be symmetric with respect to the
ecliptic plane.

{\topinsert\null\vskip-48pt
  \psfig{file=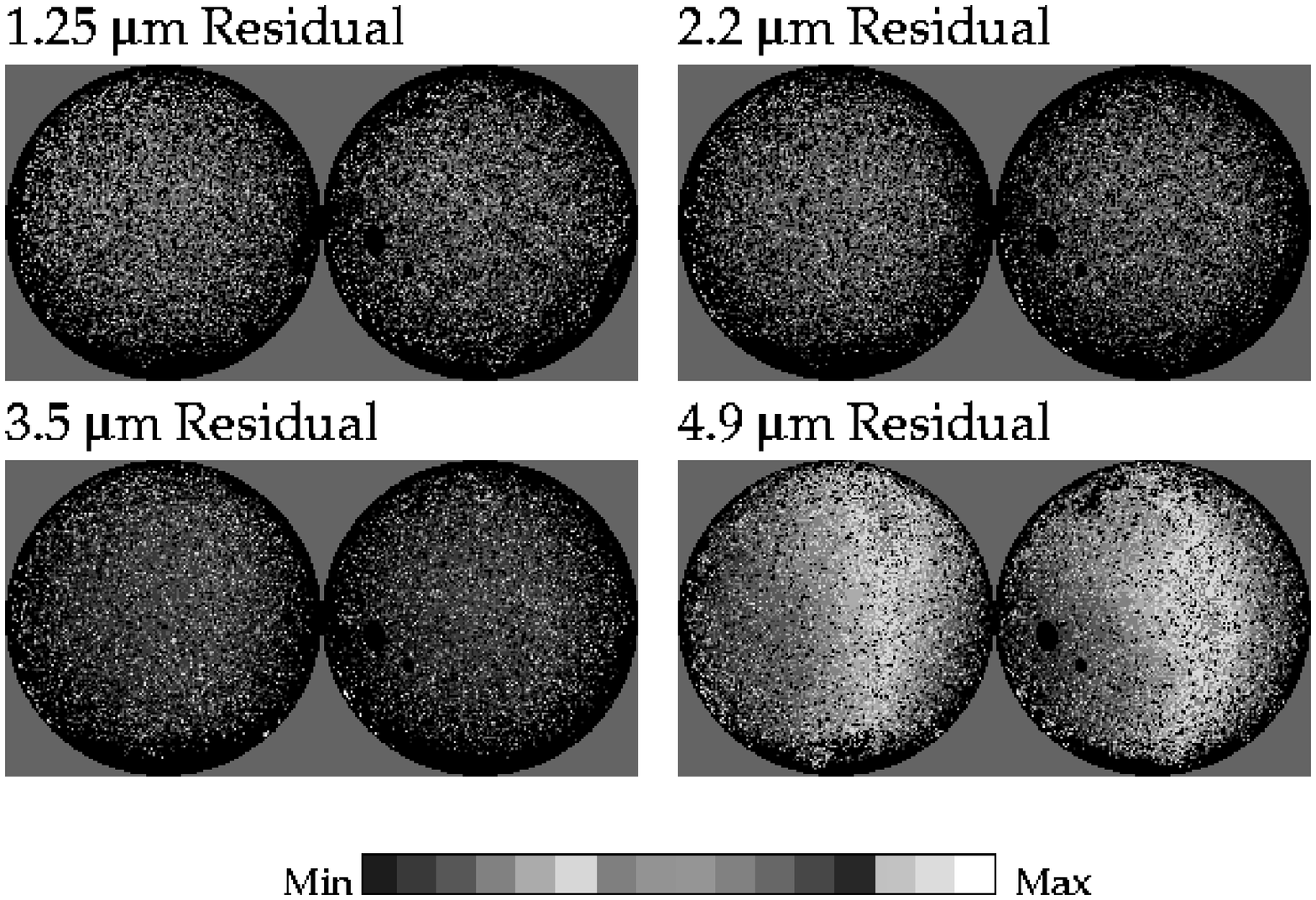,width=\h@size,angle=90,silent=yes}
  \vskip-6pt\edef\h@old{\the\hsize}\hsize=\h@size
  Fig.~{1}.---{DIRBE residual intensity maps after removal of
  foreground emission at $1.25-240~\mu$m, shown in galactic
  coordinates with an azimuthal equal-area projection.  The left
  (right) circle represents the projected north (south) Galactic
  hemisphere with $b=+90^\circ$ ($b=-90^\circ$) in the center and
  $b=0^\circ$ at the edge.  Contours of fixed latitude are concentric
  circles with $r\propto[(1-\sin|b|)/2]$.  Longitude lines run
  radially from the pole to the edge and increase clockwise
  (counterclockwise) on the left (right) hemisphere. The longitude
  $l=0\arcdeg$ runs from the center to the bottom edge of each
  projected hemisphere.  All maps are plotted on a linear scale with
  color-coded brightness ranges of $(-0.05,0.3)$, $(-0.05,0.3)$,
  $(-0.01, 0.2)$, $(0,0.2)$, $(0,2)$, $(0.5,3)$, $(0, 3)$, $(0,15)$,
  $(0,20)$, and $(0,20)$ in units of MJy~sr$^{-1}$.  Values below
  (above) the plot range are shown in black (white).}
  \vskip24pt\vskip-5.23433pt
  \hsize=\h@old\endinsert\vfuzz=0.5pt}

Analytical forms are assumed for the density distributions, scattering
phase function, and thermal emission characteristics of the dust.
Parameters for the analytical functions are determined by optimizing
the model to match the observed temporal variations in brightness
toward a grid of directions over the sky.  By fitting only the
observed time variation to determine the model parameters, Galactic
and extragalactic components of the measured brightness are totally
excluded.  However, it must be emphasized that this method can not
uniquely determine the true IPD signal; in particular, an arbitrary
isotropic component could be added to the model without affecting the
parameter values determined in our fitting to the seasonal variation
of the signal.  No such arbitrary constants are added to the
brightnesses obtained directly from our model, and limits on unmodeled
isotropic components of the IPD cloud emission are set based upon
independent knowledge of the nature of the cloud (\S~4.1).  Once the
optimal model parameters are determined, the IPD model is integrated
along the line of sight to evaluate $Z$ at the mean time of
observation of each DIRBE pixel for each week of the mission.  The
calculated IPD map is then subtracted from each DIRBE Weekly Map and
an average mission residual computed.  This simple model represents
the IPD signal fairly well, but there are clearly systematic artifacts
in the residuals at the level of a few percent of the IPD model
brightness (Paper II).  Because the zodiacal emission is so bright,
uncertainties in the residual sky maps at $12-60~\mu$m are dominated
by the uncertainties in the IPD signal.

{\topinsert\null\vskip-48pt
  \psfig{file=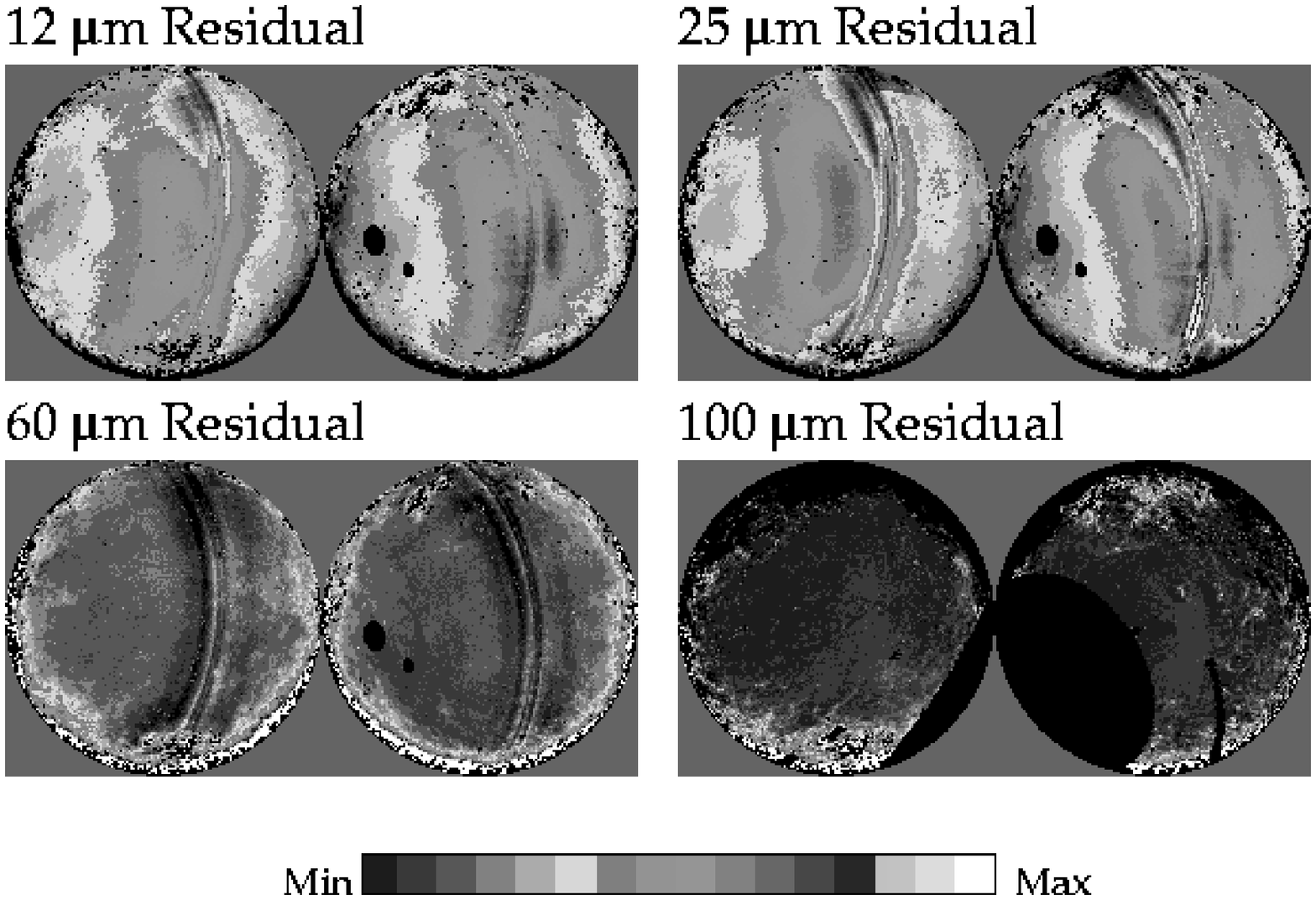,width=\h@size,angle=90,silent=yes}
  \vskip-6pt\edef\h@old{\the\hsize}\hsize=\h@size
  Fig.~{1}.---{continued.}
  \vskip24pt\vskip-5.23433pt
  \hsize=\h@old\endinsert\vfuzz=0.5pt}

The Galactic model $G$ is removed from the mission-averaged residuals
formed after removal of the IPD contribution (Paper III).  The
Galactic model actually consists of three separate components: bright
discrete sources, faint discrete sources, and the interstellar medium.
Both stellar and extended discrete sources whose intensity above the
local background exceeded a wavelength-dependent threshold are
excluded by blanking a small surrounding region from each of the ten
maps.  The blanked regions appear black in Figure 1, and are most
evident in the $1.25-4.9~\mu$m maps and at low galactic latitude.  The
contribution from faint discrete sources sources below the
bright-source blanking threshold at $1.25-25~\mu$m is then removed by
subtracting the integrated light from a statistical source-count model
based on that of Wainscoat et al. (1992), with elaborations by Cohen
(1993, 1994, 1995).  We call this the faint source model (FSM).  The
use of a source-count based model ensures that the related intensity
represents only Galactic sources.  The stellar contribution is
neglected at wavelengths longward of 25~$\mu$m.

{\topinsert\null\vskip-48pt
  \psfig{file=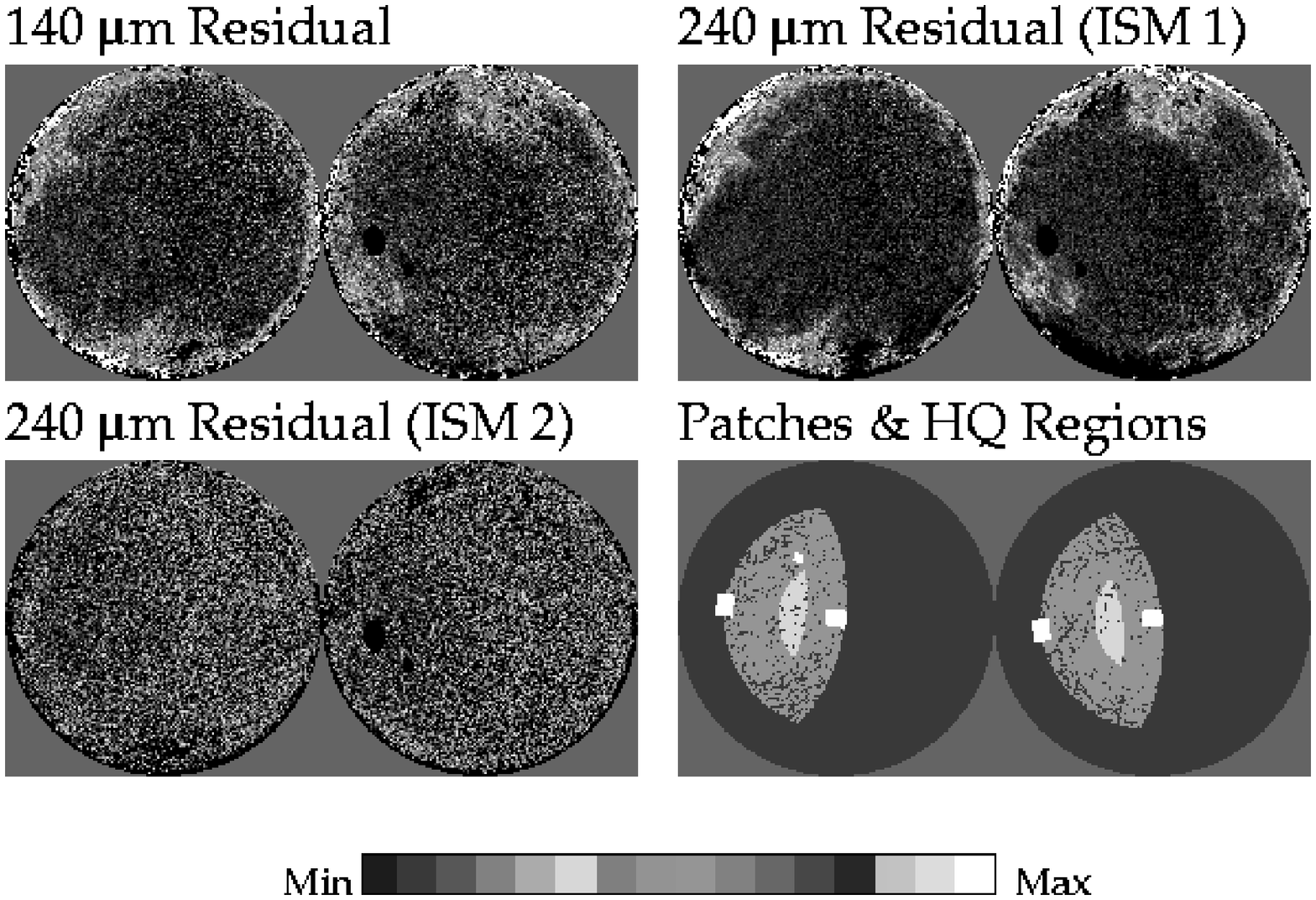,width=\h@size,angle=90,silent=yes}
  \vskip-6pt\edef\h@old{\the\hsize}\hsize=\h@size
  Fig.~{1}.---{continued. The last panel
  indicates the sky locations of the five small patches (\S~3.2) and
  the two selected high quality regions (\S~3.3): the tiny white
  square in the left hemisphere is the Lockman hole, the centered
  white square in the left (right) hemisphere is the north (south)
  Galactic pole, and the remaining white square in the left (right)
  hemisphere is the north (south) ecliptic pole; the large grey area
  in the left (right) sphere is the north (south) high quality region
  A, and the small light grey area in the left (right) hemisphere is the
  north (south) high quality region~B.}
  \vskip24pt\vskip-5.23433pt
  \hsize=\h@old\endinsert\vfuzz=0.5pt}

The basic model of emission from interstellar dust, $G_{I}(l,b,\lambda
)$, consists of a standard spatial (wavelength-independent) template
of the brightness of the interstellar medium (ISM), scaled by a single
factor $R(\lambda)$ at each wavelength.  The factor $R(\lambda)$ is
determined by the slope of a linear correlation of the standard
spatial template with the intermediate residual map at wavelength
$\lambda$ obtained from the measured map, $I_{obs}(l,b,\lambda,t)$, by
subtraction of the IPD model, blanking of bright sources, and
subtraction of the FSM.  The ISM spatial template is constructed so
that it does not contain diffuse extragalactic emission.  To the
extent that this is successful, when the scaled ISM template at any
wavelength, $G_{I}(l,b,\lambda)$, is subtracted from the intermediate
residual map at that wavelength, any CIB signal in the resulting final
residual map $I_{res}(l,b,\lambda)$ is not modified.  This linear ISM
model works well in that it removes the evident cirrus clouds,
especially in the high galactic latitude regions where the search for
the CIB is conducted.

Several approaches have been used to create the ISM spatial template.
In one approach, the 100 $\mu$m ISM map, $G_{I}(l,b,100~\mu{\rm m})$,
obtained by subtracting the contributions from the IPD and bright and
faint discrete Galactic sources from the observed map at 100 $\mu$m,
was used as the spatial template for all other wavelengths from 12 to
240 $\mu$m.  No significant ISM emission could be identified at 1.25
and 2.2 $\mu$m, and a modified form of this procedure was required to
detect the weak ISM emission at 3.5 and 4.9 $\mu$m (Paper III).  The
use of the 100 $\mu$m ISM emission as the template at other
wavelengths has the advantages of good signal-to-noise ratio and an
ideal match of angular resolution with the other DIRBE data.
Furthermore, the use of an infrared map as the template automatically
includes contributions from dust in all gas phases of the ISM.  The
procedure used to estimate the 100 $\mu$m CIB signal so as to remove
it from the 100 $\mu$m ISM map is described briefly below and in
\S~3.4.
 
For additional analysis of the 240 $\mu$m map, a two-component model
of the ISM emission (``ISM2'') was also generated.  This model used a
linear combination of the DIRBE 100 and 140 $\micron$ ISM maps as a
template.  While the one-component (100 $\mu$m) model (``ISM1'')
appears to work adequately at high latitudes, where we could best test
for isotropic residuals, the ISM2 model can account for spatial
variations in dust temperature throughout the ISM (Paper III).  This
leads to a more accurate model of the ISM emission, particularly at
low galactic latitudes, and a residual map $I_{res}(l,b,240~\mu{\rm
m})$ that is more weakly correlated with the ISM template than in the
case of the ISM1 model.  Figure 1 shows maps of $I_{res}(l,b,\lambda)$
at 240 $\mu$m for both the ISM1 and ISM2 models.

To search for evidence of an isotropic CIB residual at 100 $\mu$m, an
ISM spatial template independent of the measured 100 $\mu$m map was
needed.  For this purpose a velocity-integrated map of H~I column
density was used as the spatial template of the ISM emission.  The
range of velocities in the H~I map was restricted so that it contained
only Galactic H~I emission.  The success of this procedure of course
depends on the accuracy with which the H~I traces the dust
distribution, at least at the high galactic latitudes of interest
here.  Paper III provides extensive discussion of the uncertainty in
the correlation of infrared brightness with H~I column density.

The H~I spatial template used to remove ISM emission from the map at
100 $\mu$m was the Bell Labs H~I survey (Stark et al. 1992).  This
survey has the advantages of a well-established baseline and large
area coverage, but the disadvantage of lower angular resolution than
the DIRBE data.  Higher resolution H~I data (Elvis, Lockman, \&
Fassnacht 1994; Snowden et al. 1994), obtained in small regions where
there are observational constraints on the amount of molecular and
ionized material (Paper III), and calibrated with the Bell Labs H~I
survey, were used to establish the scaling factor between the H~I and
100 $\mu$m ISM emission.  These same high resolution data were used to
estimate the 100 $\mu$m brightness at zero H~I column density so as to
remove diffuse extragalactic emission from the ISM spatial template,
$G_{I}(l,b,100~\mu{\rm m})$, used at all other wavelengths as
discussed above (see \S~3.4).

The 100 $\mu{\rm m}-{\rm H~I}$ correlation was also evaluated using
the new Leiden/Dwingeloo H~I survey (Hartmann \& Burton 1997), but
this made little difference in the scaling factor or the residual
intensity $I_{res}(l,b,100~\mu{\rm m})$.  Use of the Leiden/Dwingeloo
H~I survey as the spatial template of the ISM at 100 $\micron$
produces a cleaner map of residual emission $I_{res}(l,b,100~\mu{\rm
m})$ than does use of the Bell Labs data, because of a better match to
the DIRBE angular resolution, but the differences are not very
apparent in maps made in the projection and scale of those in Figure
1. Results quoted in this paper are based on the Bell Labs H~I survey
and other observations that are directly calibrated to that data set
(Elvis, Lockman, \& Fassnacht 1994; Snowden et al. 1994).

\subsection{2.4. Uncertainties}

For this analysis it is useful to make distinctions between three
forms of uncertainties.  First are the {\it random uncertainties}
which include instrumental noise, uncorrected instrument gain
variations, random fluctuations of the stellar distribution, and
certain deficiencies in the foreground modelling procedures.  The key
property of random uncertainties is that they are reduced as one
averages over longer time intervals or larger regions of the sky.
Table 2 lists typical values for the detector noise per pixel averaged
over the entire mission, $\sigma(\nu I_\nu)$, assuming 400
observations per pixel.  The bolometer detectors used at 140 and 240
$\mu$m are distinctly less sensitive than the other detectors.

The second form of uncertainty is the {\it gain uncertainty}.  This is
the uncertainty in the gain factor used in the absolute calibration of
the DIRBE data.  While the gain uncertainty does affect the quoted
intensities, including the residual intensities, in a systematic way,
it does not alter the signal-to-noise ratio of the results or the
detectability of an isotropic residual signal using our methods.  We
therefore distinguish the gain uncertainty, shown as $S({\rm gain})$
for each wavelength band in Table 2, from other systematic errors.

Finally there are the {\it systematic uncertainties}, which are the
uncertainties in the data and the foreground models that tend to be
isotropic or very large scale.  The systematic uncertainties cannot be
reduced by averaging, and therefore are the ultimate limitations in
the detection of the CIB.  Table 2 lists the detector offset
uncertainties, $\sigma({\rm offset})$. The offset uncertainties are
important contributors to the total uncertainty only at $140$ and
$240~\mu$m.  The systematic uncertainties of the IPD model, the
stellar emission model, and the ISM model are important respectively
at $1.25-100~\mu$m, $1.25-4.9~\mu$m, and $100-240~\mu$m.  Papers II
and III discuss in detail the estimation of the systematic
uncertainties in the foreground models; Table 6 of Paper III lists the
systematic uncertainty associated with each foreground.  The
systematic uncertainty in each residual shown in Table 2 of this paper
is the quadrature sum of the individual contributions identified in
Paper III.  The total uncertainties used to state our most restrictive
upper limits on the CIB, and the uncertainty in the CIB detections at
140 and 240 $\mu$m, are estimated as the quadrature sum of the random
and systematic uncertainties.  Table 2 of this paper and Table 6 of
Paper III clearly show that {\it the total uncertainties are dominated
by the systematic uncertainties in removing the foreground
contributions to the infrared sky brightness}.

\section{3. OBSERVATIONAL RESULTS}

\subsection{3.1. Dark Sky Limits}

The most conservative direct observational limits on the CIB are
derived from the minimum observed sky brightnesses.  In each DIRBE
weekly sky map, the faintest direction has been determined for each
wavelength.  At wavelengths where interplanetary dust scattering or
emission is strong, the sky is darkest near the ecliptic poles.  At
wavelengths where the IPD signal is rather weak (i.e., longward of 100
$\mu$m), the sky is darkest near the galactic poles or in minima of
H~I column density. The smallest of these values at each wavelength
over the duration of the mission is the ``dark sky'' value, listed in
Table 2 as $\nu I_\nu({\rm dark})$.  The uncertainty shown for each
value is the quadrature sum of the contributions from the gain and
offset 1$\sigma$ uncertainties.  We define ``dark sky'' upper limits
to the CIB at the 95\% confidence level (CL) as $2\sigma$ above the
measured dark sky values.

\subsection{3.2. Residuals in Small Dark Patches}

After removing the contributions of interplanetary dust (Paper II),
\hskip-0.99pt
bright and faint discrete galactic sources, and the interstellar
medium (Paper III) from the measured sky brightness, the residual
signal at high galactic and ecliptic latitudes is positive and
generally rather featureless, though low level artifacts from
systematic errors in the models are clearly present.  To illustrate
the magnitude of the foreground signals, Figure 2 shows the DIRBE
spectrum of the total observed sky brightness averaged over a
$5^\circ\times5^\circ$ region at the Lockman hole, the region of
minimum H~I column density at $(l,b)\sim(150^\circ,+53^\circ)$
[geocentric ecliptic coordinates $(\lambda,\beta)\sim(137^\circ,
+45^\circ)]$ (Lockman, Jahoda, \& McCammon 1986; Jahoda, Lockman, \&
McCammon 1990).  Figure 2 also shows the individual contributions from
the foreground sources and the residuals after removing the foreground
contributions.  Scattering and emission from the interplanetary dust
dominates all other signals from 1.25 to 100 $\mu$m.  This is true
even at 3.5 $\mu$m, the spectral ``window'' between the maxima of the
scattered and emitted IPD signal.  Only at 140 and 240 $\mu$m does
some other foreground signal, that from the interstellar medium
(infrared cirrus), become dominant.

Some insight into the residuals is provided by looking at several high
latitude regions (Hauser 1996a, b).  For this purpose, we have
examined the residuals in $10^\circ\times10^\circ$ fields at the north
and south Galactic and ecliptic poles (designated NGP, SGP, NEP, and
SEP respectively), and a $5^\circ\times5^\circ$ field in the Lockman
hole (LH).  Table 2 lists the mean residual brightnesses for these
five patches after all of the foreground removal steps listed above.
As discussed in \S~2.3, the 100~$\mu$m map was used as the ISM
template in producing the residual maps at all wavelengths except
100~$\mu$m.  At $100~\mu$m, the Bell Labs H~I map was used as the ISM
template.  While the range of residual values at each wavelength is
substantial, typically a factor of 2 or more, comparison with the dark
sky values shows that these residuals are small fractions, approaching
10\% at wavelengths shortward of 100 $\mu$m, of the dark sky values.
However, the fact that the residuals are brightest in the region of
peak IPD thermal emission, 12 to 25 $\mu$m, strongly suggests that
significant foreground emission still remains, at least in the middle
of the DIRBE spectral range.  This is not surprising in view of the
very apparent residual IPD modelling errors at these wavelengths
(e.g., Fig. 1, especially 4.9 to 100 $\mu$m; and Paper II).

\subsection{3.3. Residuals in High Quality Regions}

While each of the small dark patches (\S~3.2) is situated where one of
the IPD, stellar, or ISM foregrounds is minimized, each patch is also
located in a region where the other foregrounds may be strong.
Therefore, we defined ``high quality'' (HQ) regions where all
foregrounds are expected to be relatively weak.  The range of ecliptic
latitude, $\beta$, was restricted to exclude bright scattering and
emission from the IPD, and the range of galactic latitude, $b$, was
restricted to exclude regions with bright stellar emission.  To avoid
regions with bright ISM emission, locations where the 100 $\mu$m
brightness, after the IPD contribution was removed, was more than 0.2
MJy~sr$^{-1}$ above the local mean level were also excluded.  The
largest region that can reasonably be considered as high quality
covers $\sim20$\% of the sky between the Galactic and ecliptic poles
and is designated HQA.  A much more restrictive region, designated
HQB, lies in the center of the HQA region and includes $\sim2$\% of
the sky.  Table 3 lists the constraints for the HQ regions, and the
last panel of Figure 1 shows the areas covered by the HQ regions.
Each HQ region is composed of corresponding northern and southern
segments.

Table 2 lists the mean residual intensities, $\nu I_0({\rm HQA})$ and
$\nu I_0({\rm HQB})$, for the HQ regions after all foregrounds have
been removed.  As in the analysis of the small dark patches (\S~3.2),
the 100~$\mu$m map was used as the ISM template in producing the
residual maps at all wavelengths except 100 $\mu$m.  At $100~\mu$m,
the Bell Labs H~I map was used as the ISM template.  The statistical
uncertainty of the mean, which is calculated from the observed rms
variation of the residual emission over the region, is also shown. For
HQB, the total systematic uncertainty estimated for each band is also
listed in Table 2.  While some portions of the systematic uncertainty
needed to be evaluated at regions other than the HQ regions (see
Papers II and III), the numbers listed here should be appropriate for
HQB.  The systematic uncertainties are larger when dealing with other
areas where the foreground emission removed was stronger.

{\topinsert\null\vskip-12pt
  \psfig{file=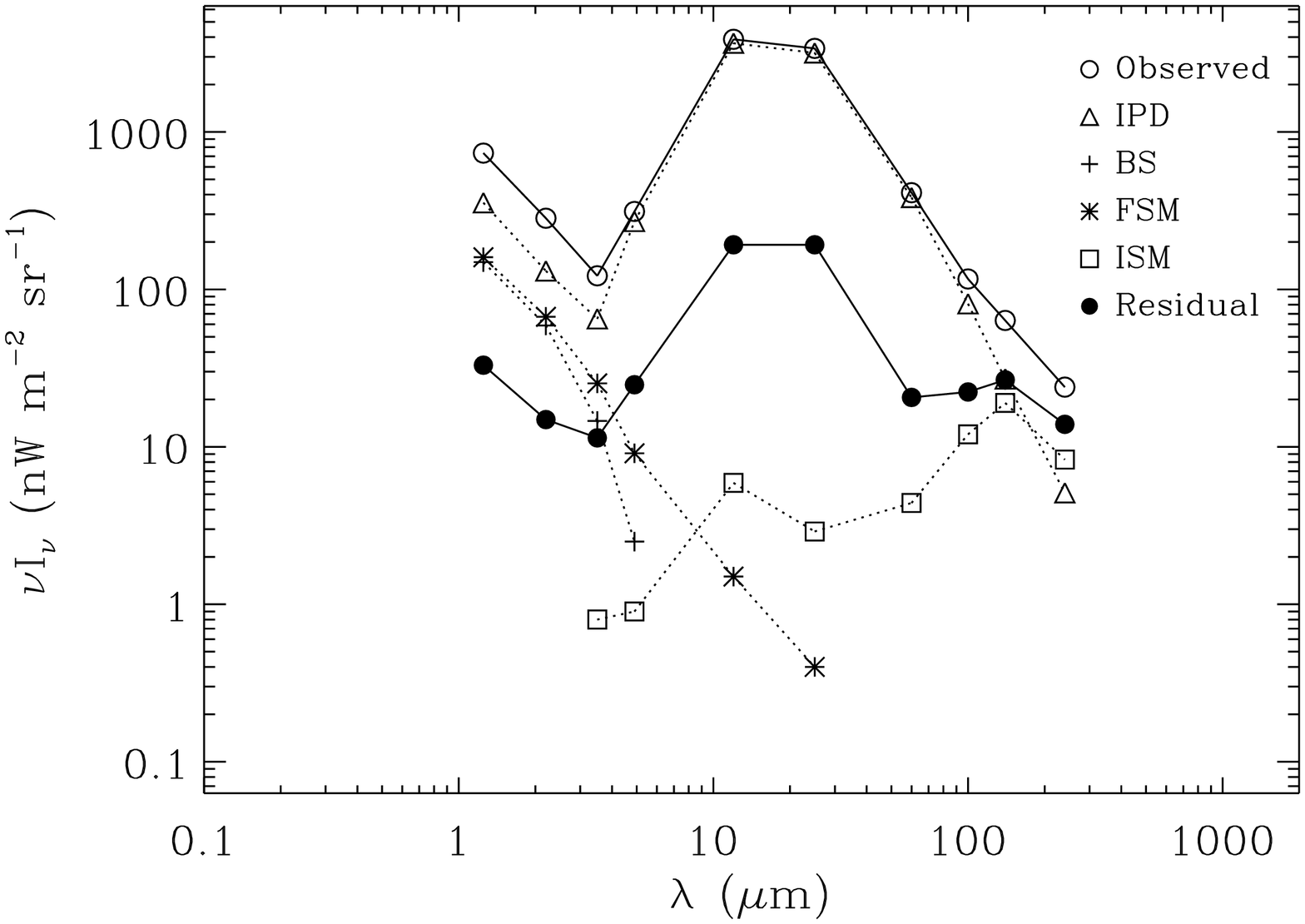,width=\h@size,angle=90,silent=yes}
  \vskip12pt\edef\h@old{\the\hsize}\hsize=\h@size
  Fig.~{2}.---{Contributions of foreground emission to the DIRBE data
  at $1.25-240~\mu$m in the Lockman hole area: observed sky brightness
  (open circles), interplanetary dust (triangles), bright galactic sources
  (crosses), faint galactic sources (stars), and the interstellar
  medium (squares).  Filled circles show the residual brightness after 
  removing all foregrounds from the measurements.}
  \vskip24pt\vskip-5.23433pt
  \hsize=\h@old\endinsert\vfuzz=0.5pt}

\subsection{3.4. Residuals at the Lockman Hole}
\vskip-12pt
\subsection{and the North Ecliptic Pole}

The intercept of a linear fit to the correlation between the infrared
emission and the H~I column density yields an estimate of the
isotropic residual component of infrared emission.  This technique was
used (\S~2.3 and Paper III) to establish the amount of emission that
needed to be removed to create the 100 $\micron$ template of the ISM.
The H~I data were from Snowden et al. (1994) for a $250$~deg$^2$
region covering the Lockman hole (LH$^\prime$), and from Elvis,
Lockman, \& Fassnacht (1994) for a $70$~deg$^2$ region around the
north ecliptic pole (NEP$^\prime$).  These regions are denoted with
primes to distinguish them from the ``dark patches'' LH ($5^\circ
\times 5^\circ$ patch) and NEP($10^\circ\times10^\circ$ patch) at
similar locations but of somewhat different size discussed in \S~3.2.
Figures 7 and 8 of Paper III show that the 100 $\mu$m brightness and
H~I column density are linearly related at low column density in these
regions.  Within these regions, linear fits to the correlations
between the 140 and 240 $\mu$m emission and the H~I column density
were also calculated.  Table~2 lists the intercepts of these fits as
$\nu I_0({\rm LH^\prime,HI})$ and $\nu I_0({\rm NEP^\prime,HI})$.

The advantage of this technique for estimating the CIB at 140 and 240
$\mu$m, over our standard method using the 100 $\mu$m data for the ISM
template, is that the systematic uncertainties of the 100 $\mu$m data,
including those caused by uncertainties in the 100 $\mu$m IPD model
and in the extrapolation of the 100 $\mu$m---H~I correlation to zero
H~I column density, are not propagated into the 140 and 240 $\mu$m
results.  Thus, the systematic uncertainties for $\nu I_0({\rm
LH^\prime,HI})$ are smaller than those for $\nu I_0({\rm HQB})$. For
the NEP region, the intercept of the correlation must be extrapolated
over a longer interval of H~I column density and from fewer data, so
the systematic uncertainties for $\nu I_0({\rm NEP^\prime,HI})$ are
only smaller than those of $\nu I_0({\rm HQB})$ at 100 $\mu$m.

A disadvantage of this technique is that the H~I does not trace other
phases of the ISM (ionized and molecular gas) that may also contribute
to the observed infrared emission.  Any part of the emission from
other phases that is not directly correlated with the H~I column
density will appear as an additional contribution to $\nu I_0({\rm
LH^\prime,HI})$ and $\nu I_0({\rm NEP^\prime,HI})$.  Additionally,
even within the neutral ISM, the assumed linear correlation between
infrared brightness and H~I column density cannot track large- or
small-scale variations in the dust temperature or gas-to-dust mass
ratio.  This is apparent in the 100 $\mu$m residual map (Fig.~1),
where the ISM emission is strongly oversubtracted in the outer Galaxy
and undersubtracted in the inner Galaxy.  Emission from numerous
molecular clouds is also visible at high latitudes.

Because there are available data on all gas phases in the NEP$^\prime$
and LH$^\prime$ regions, it is possible to set tight limits on the
uncertainty in the extrapolation of the infrared---H~I correlation to
zero H~I column density in these regions (Paper III).  We estimate
that dust in the ionized ISM uncorrelated with H~I contributes less
than $4\nWm2sr$ to the 100 $\micron$ residual intensity.  Assuming
that the infrared spectrum of the ionized medium is the same as that
of the neutral medium, the contributions from the ionized ISM at 140
and 240 $\micron$ are less than $5\nWm2sr$ and $2\nWm2sr$
respectively.  If such large contributions were to exist, then the
residual intensities listed in Table 2 would have to be reduced
accordingly.  Even in this case, the 240 $\micron$ result would still
be a 3 $\sigma$ detection of residual emission.

Analysis in Paper III also shows that infrared emission from the
molecular ISM is only poorly constrained by upper limits on CO
observations.  Constraints based on visual extinction measurements
suggest the contribution from dust in the molecular ISM is negligible
at 100 $\micron$.  Contributions at 140 and 240 $\micron$ should be
similarly low.

\footnote{}
{\eightpoint\vskip0pt
 \tabbot{3}{H{\sc IGH} Q{\sc UALITY} R{\sc EGION} D{\sc DEFINITIONS}}
 {
  \hfil#\hfil&\hfil#\hfil&\hfil#\hfil&\hfil#\hfil&
  \hfil#\hfil&\hfil#\hfil&\hfil#\hfil\cr
  \tabhead{
   {}&{}&{}&{100 $\micron$ ISM}&{}&{Area$^a$}&{}\cr
   \noalign{\vskip-3pt\hskip4.3in${\hskip2in\over\hskip2in}$\vskip-3pt}
   {Region}&{$|b|$ limit}&{$|\beta|$ limit}&{limit (MJy/sr)}&{(pixels)}&
   {(deg$^2$)}&{(sr)}\cr}
  \tabbody{
HQA & $>30$ & $>25$ & $<0.2$ & 83671 & 8780 & 2.67\nl
HQB & $>60$ & $>45$ & $<0.2$ & \hskip4pt8140 & \hskip4pt854 & 0.26\nl}
 }
 \tabnote{$^a$ Bright source removal reduces these areas by up
to 35\% at near-IR wavelengths.}
\vskip6pt
}

\subsection{3.5. Isotropy of the Residual Emission}

The signature of the diffuse CIB is an isotropic signal. Several tests
of the isotropy of our residual signals have therefore been
performed. Fundamentally, each test checks whether the background
intensities in different directions agree within the limit of the
estimated uncertainties.

\subsection{3.5.1. {\it Mean Patch Brightnesses}}

The first test involves comparison of the mean brightnesses of the
small dark patches discussed in \S~3.2. For each patch the mean
brightness and the standard deviation of the mean ($\nu I_0\pm
\sigma_m$) are listed as the residual value and random error in Table
2.  Two patches whose means differ by less than 2$\sigma_m({\rm
total})=2\sqrt{\sigma_m(1)^2+\sigma_m(2)^2}$ are consistent with
isotropy between those regions of the sky. This is a strict constraint
on isotropy, in that it does not allow for differences between patches
that are larger than the random errors but within the systematic
uncertainties.

Some pairs of patches pass this strict test for isotropy at 1.25, 2.2,
3.5, 4.9, 140 and 240 $\mu$m.  For the ISM2 model (\S~2.3), the mean
240 $\mu$m residual intensity of each patch except the NEP is
consistent with that of each of the other patches.  However, in most
cases the differences between the mean residuals of the patches are
larger than expected for purely random noise in measurements of an
isotropic residual.  At mid-infrared wavelengths, the systematic
effect of the residual IPD emission is evident in that the north and
south ecliptic pole patches are at nearly the same brightness, while
the lower ecliptic latitude patches at the Galactic poles are
significantly brighter.

If the criterion for isotropy is taken to be agreement within the
systematic uncertainties, which are also shown in Table 2, then most
pairs of patches pass the test at all wavelengths.  Exceptions are
that intensities at the Galactic poles tend to differ from those at
the ecliptic poles at wavelengths where the IPD emission is strong,
and the residual intensity in the SEP patch is anomalously low in the
near-infrared and high in the far-infrared.

A test for equal mean intensities was also applied for the north and
south halves of the HQB region.  In this case, the confidence levels
of the equality were determined through the bootstrap method and the
$t$-statistic of the Fisher-Behrens test:
$$
t={{\overline{N}-\overline{S}}\over{\sqrt{\sigma_N^2/n_N+\sigma_S^2/n_S}}},
\eqno(2)
$$
where $\overline{N}$ and $\overline{S}$ are the mean intensities over
$n_N$ and $n_S$ pixels in the north and south halves of the HQB
region.  Only at 3.5 $\mu$m and 240 $\mu$m were the two means
plausibly equal, to significance levels of 36\% and 75\% respectively.
At the other wavelengths, the highest significance level of equality
was only 0.3\% (at 140 $\mu$m).

\subsection{3.5.2. {\it Brightness Distributions}}

The next set of isotropy tests involves checking whether the
dispersion in brightness for pixels in an area is consistent with the
dispersion due to the known random uncertainties. If the data show no
variation in excess of that expected from the random uncertainties
then the patch is said to be isotropic. This test gains statistical
significance when large patches are used. We applied this test in the
HQ regions defined in \S~3.3.

For wavelengths of $12-240~\mu$m, the probability distributions for
the intensity of each pixel were calculated assuming Gaussian
dispersions of both ISM model errors (proportional to the ISM
intensity) and a combination of detector noise and IPD model errors
(measured at each pixel as the standard deviation of the weekly map
intensities, after removal of IPD emission). The random uncertainties
of the stellar model are included as an additional Gaussian component
to the dispersion at 12 and 25 $\mu$m, even though the contribution
from stars is small enough that this additional term is minor.  The
expected intensity distribution for the entire patch is then
constructed from the sum of these Gaussian distributions over all
pixels.

For wavelengths of $1.25-4.9~\mu$m, the residual fluctuations from
faint sources dominate the variations within the HQ regions.  For
these wavelengths and for each HQ region, the faint source model
(\S~2.3) was used to generate random samples of pixel brightnesses
using Poisson statistics. We then added random Gaussian errors
corresponding to the combined detector noise and IPD model
uncertainties, and the ISM uncertainties at wavelengths for which the
ISM was modeled (3.5 and 4.9 $\mu$m).

For all wavelengths, the observed and expected residual brightness
distributions were compared using the Kolmogorov-Smirnov (K-S) test.
At wavelengths greater than $12~\mu$m the $\chi^2$ test was also
applied.  These statistics indicate isotropy for the 240 $\micron$
residuals in the HQA (ISM1) and HQB (ISM1 \& ISM2) regions. The
residuals in the HQB region are also found to be isotropic at 140
$\micron$. The 60 and 100 $\micron$ intensity distributions fail the
tests, despite their qualitatively similar observed and expected
distributions. The 12 and 25 $\micron$ distributions fail the test
badly, because of residual structure from imperfect removal of the IPD
emission. At the near-infrared wavelengths, the FSM predicts slightly
wider distributions than are observed.

This brightness distribution test moves beyond the simple comparison
of mean intensities and can reveal the presence of unusually bright or
dark features within a region. The main drawback of this test is that
it lacks any sensitivity to the spatial distribution of the residual
emission within a region.

\subsection{3.5.3. {\it Systematic Spatial Variations}}

An area where the residual intensity is isotropic will have no
significant spatial variations or structure. The residual emission in
the HQ regions has been tested for systematic variations by looking
for linear correlations of the residual intensity with $\csc(|b|)$ and
$\csc(|\beta|)$, and with the intensities of the IPD, faint source and
ISM models. The slopes of these correlations indicate the gradients in
the residual intensities with respect to each correlant. There were
statistically significant slopes to all of these correlations in the
HQA regions. These correlants are not all independent. Correlations
with all of them can be produced by low-level artifacts due to
imperfections in any one of the foreground models. Examination of the
residual maps (Fig.~1) shows evident residuals from the IPD and ISM
model removal in the HQA region, consistent with these formal tests.
In the HQB regions, the residual emission at 140 and 240 $\micron$ did
not exhibit any significant correlations, even though the tests were
sensitive enough to detect correlations as strong as those found in
the larger HQA regions. At other wavelengths, correlations with at
least one of the models were present. For HQB, the slopes of the
correlations with respect to $\csc(|b|)$ and $\csc(|\beta|)$ and their
statistical uncertainties are listed as $\partial_b\nu I_0({\rm HQB})$
and $\partial_{\beta}\nu I_0({\rm HQB})$ in Table 2. The high quality
regions HQA and HQB were defined {\it a priori} as regions of least
solar system and galactic foreground, not based upon the outcome of
isotropy tests of residuals.  Since the HQA region contains evident
model artifacts, the remaining tests were restricted to the HQB
region.

A more general test for structure within an area, such as the HQB
region, is to fit a trend surface. If the scatter of the residuals of
the fit is significantly less than the scatter about the mean value in
the patch, structure exists. To determine the significance of a
measure of scatter in a patch, without making any assumptions about
the nature of the data, the intensity values of its pixels were
randomly permuted spatially, creating a ``flat'' reference
patch. Applying a surface fit to many such randomized versions of a
patch allowed the derivation of the empirical distribution function of
the $\chi^2$ of the fit to a flat patch. The same type of surface was
then fitted to the actual patch data (no permutation), and the
$\chi^2$ calculated.  The fraction of randomized patches with smaller
values of $\chi^2$ is the significance level to which the patch is
flat.  This analysis was performed individually on the two HQB
patches, using polynomials through degree $n$ in a galactic coordinate
system, $l$ and $\csc b$.

\footnote{}
{\eightpoint\vskip6pt
 \tabbot{4}{T{\sc EST FOR} S{\sc URFACE} T{\sc RENDS IN} HQB}
 {
  \hfil#\hfil&\hfil#\hfil&\hfil#\hfil&\hfil#\hfil&
  \hfil#\hfil&\hfil#\hfil&\hfil#\hfil\cr
  \tabhead{
   \noalign{\hskip3.1in Flatness of HQBN, HQBS (\%)}
   \noalign{\vskip-3pt\hskip1.4in${\hskip4.9in\over\hskip4.9in}$\vskip-3pt}
   {Surface}&{1.25 $\micron$}&{2.2 $\micron$}&{3.5 $\micron$}&
   {4.9 $\micron$}&{140 $\micron$}&{240 $\micron$}\cr}
  \tabbody{
$P_1(l,\csc b)$&47, 36& 50, 43 & 48, 45 & 28, 10 & 50, 52 & 51, 52 \nl
$P_2(l,\csc b)$&46, 26& 48, 29 & 38, 40 & 22, \hskip4pt6 & 50, 50 & 52, 49\nl
$P_3(l,\csc b)$&44, 24& 46, 39 & 38, 39 & 23, \hskip4pt4 & 50, 49 & 53, 50\nl}
 }
 \vskip-6pt
}

Table 4 lists the results of this analysis applied to the separate
north and south halves of the HQB region (HQBN and HQBS) for a surface
of up to degree 3 (10 terms). The entries at wavelengths of 12 to 100
$\mu$m are omitted since at these wavelengths the residuals are
clearly not isotropic: surface trends are obvious and the significance
level of flatness less than 0.1\%.  The HQBS region at 4.9 $\mu$m also
bears evidence of structure, but the test was inconclusive for the
other entries in Table~4. One can only say they are consistent with
being flat.  However, there may be clustering or some other irregular
structure which, to a smooth polynomial surface, appears as noise.

\subsection{3.5.4. {\it Two-Point Correlation Functions}}

A more sophisticated test of the isotropy of the residual infrared
emission is the two-point correlation function of the residuals. The
procedures used were very similar to those employed for the analysis
of the CMB anisotropy in the {\it COBE}/DMR data (Hinshaw et
al. 1996).  The two-point correlation function is expressed as
$C(\theta) = \langle \nu I_i\nu I_j \rangle$ where the angle brackets
denote the average over all $N_{ij}$ pixel pairs in the region of
interest that are separated by an angular distance $\theta$. The pixel
intensities, $I_i$, have had the mean residual intensity (i.e., the
monopole term) subtracted.

Figures 3a, 3b, and 3c show the two-point correlation functions for
the 3.5, 100, and 240 $\micron$ residual emission in the HQB region.
The correlation function is binned into $0\fdg25$ bins, which is
slightly less than half of the width of the DIRBE beam. The degree of
isotropy of the two-point correlation function was evaluated by
comparing the correlation function of the real data with two-point
correlation functions generated from an ensemble of Monte Carlo
simulations of the residual brightness in the HQB regions. The
simulations assumed zero mean intensities with random Gaussian
uncertainties in each pixel that were estimated from the weekly
variation of the observed data, after removal of the IPD emission (see
Paper III for details). We increased the number of simulations until
the statistical results (below) were unaffected by the size of the
sample. This required 7200 simulations at 240 $\micron$ (9600 for
ISM2), and 4800 simulations at 140 $\micron$. At other wavelengths,
comparison of the observed correlation function with the theoretical
uncertainties ($\sigma_{C(\theta)} = \sigma_I^2 / \sqrt{N_{ij}}$,
assuming that a single $\sigma_I$ applies for all pixels) was
sufficient to demonstrate a clear lack of isotropy.

For the data and each of the simulations, a $\chi^2$ statistic
was calculated as
$$
\chi^2={\bf\Delta C^T\cdot M^{-1}\cdot\Delta C} \eqno(3)
$$
where $({\bf \Delta C})_i = C(\theta_i) - \sigma^2\delta(0)$ is the
difference between the correlation function of the data (or one of the
simulations) and the correlation function for a perfectly isotropic
distribution [$C(\theta)=0$ except $C(0) = \sigma^2$] and ${\bf
M^{-1}}$ is the inverse of the correlation matrix ${\bf M = \langle
(\Delta C)(\Delta C)^T\rangle}$, where the angle brackets here denote
an average over all Monte Carlo simulations. If there were no
cross-correlation between the terms of the two-point correlation
function, this definition of $\chi^2$ would reduce to the usual
form. The lines in the two-point correlation of Figure 3c indicate the
$\pm 1\sigma$ (rms) variations in $C(\theta)$ for all the simulated
correlation functions at 240 $\micron$.  In Figures 3a and 3b, the
lines indicate the theoretically expected variation of
$\sigma^2/\sqrt{N_{ij}}$ for the 3.5 and 100 $\micron$ data.

{\topinsert\null\vskip-12pt
  \psfig{file=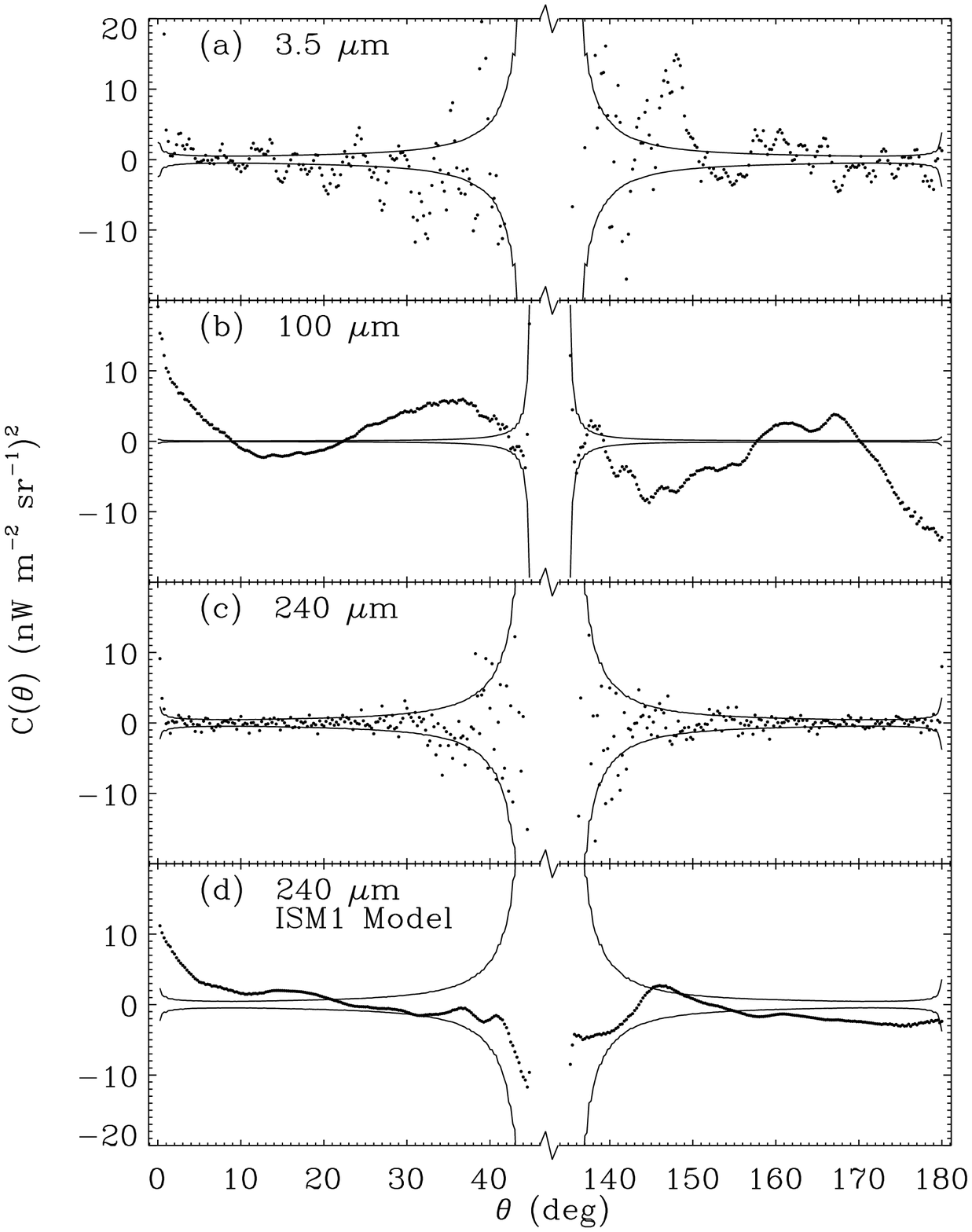,height=6.2in,width=\h@size,angle=0,silent=yes}
  \vskip12pt\edef\h@old{\the\hsize}\hsize=\h@size
  Fig.~{3}.---{Two-point correlation function used to test the isotropy
  of the DIRBE residual emission in the high quality region B (Table
  3).  The top three panels show the correlation functions of the
  residuals at three wavelengths, while the bottom panel shows the
  correlation function of the interstellar medium model (ISM1) at
  $240~\mu$m.  The solid lines in each panel are the $\pm1\sigma$
  uncertainties estimated by Monte Carlo simulations.  Separations of
  $0^\circ\leq \theta<45^\circ$ are obtained within each of the north
  and south high quality B regions, while separations of
  $135^\circ<\theta \leq180^\circ$ are obtained between the north and
  south high quality B regions. The large uncertainties at
  $\theta\approx45^\circ$ and $135^\circ$ are due to the small number
  of pixel pairs at these separations.}
  \vskip24pt\vskip-5.23433pt
  \hsize=\h@old\endinsert\vfuzz=0.5pt}

Ideally, if the data are isotropic the reduced $\chi^2$,
$\chi^2_{\nu}$, should be $\approx 1.0$ and the fraction of
simulations that have a smaller $\chi^2$ than the data should be
$P(<\chi^2) \approx 0.5$. Table 5 lists the results at 240 $\micron$
for the entire HQB region, and for the north and south halves
considered independently.  The results of the analysis of the
residuals from the two-component ISM model (ISM2) are also presented.
Within the subsets of the HQBS (ISM1) and HQBN (ISM2) regions, the 240
$\micron$ data are found to be indistinguishable from the random
simulations.  In the full HQB region, the $P(<\chi^2)$ values, while
more marginal, do not support rejection of the hypothesis that the
residual 240 $\micron$ emission in the HQB region is isotropic.  As a
further comparison, Figure 3d shows the correlation function in the
HQB region for the ISM1 model used in creating the 240 $\mu$m residual
map.  Structure of this character is absent in the 240 $\mu$m residual
map (Fig.~3c).  The 100 $\mu$m ISM map was clearly a good template for
the 240 $\mu$m ISM emission.  On the other hand, the large features in
the correlation function of the 100 $\mu$m residual map (Fig.~3b),
which was created using an H~I map as the ISM template, indicate that
there generally are some deficiencies in the assumption that H~I is an
accurate spatial tracer of dust (as noted in \S~3.4).

For the 140 $\micron$ residual emission, the case for isotropy of the
residual emission is not as strong, but is still not thoroughly
rejected (Table~5).  At wavelengths shorter than 140 $\mu$m, isotropy
can be ruled out by the fact that the two-point correlation functions
display significant structure (caused by imperfect removal of
foreground emission) and $\chi^2_{\nu} \gg 1.0$ (e.g., Figs.~3a
and~3b).

As a further check on the isotropy of the residual emission in the HQB
region, we also calculated the two-point cross-correlations between
the residual 140 $\micron$ emission and the IPD and ISM models used in
deriving those residuals. The same cross-correlations were calculated
for both the ISM1 and ISM2 residual emission maps at 240 $\micron$.
Table 5 shows the results of these cross-correlations. The
cross-correlations indicate isotropy at about the same level of
confidence as the auto-correlations.

Table 5 also includes the results found when the two-point correlation
functions of the residual emission at 140 and 240 $\micron$ are
calculated over the region of the Lockman hole. For this test, the
residual emission was generated by the subtraction of the H~I emission
scaled by the slope of the H~I - IR correlation (\S~3.4), rather than
the standard 100 $\micron$ template of the ISM emission. The LH region
is smaller than the HQB region, and only samples angular separations
in the range $0\arcdeg\leq\theta\lesssim22\arcdeg$.  Using the H~I
column density as the ISM template, the 140 $\micron$ residual
emission in the LH region exhibits isotropy at roughly the same level
of confidence as does the 140 $\micron$ residual in the HQB region
when the 100 $\micron$ data are used as the ISM template. At 240
$\micron$, use of the H~I data as the ISM template leads to residuals
that are indistinguishable from the random simulations. Apparently,
within this region of low H~I column density, there is little or no
indication of anisotropic emission from the ionized and molecular
phases of the ISM.

Two-point correlation functions for the 140 and 240 $\micron$ residual
emission often exhibit an increase at the smallest angular scales,
$\theta < 1\arcdeg$ (e.g., Fig.~3c).  However, any apparent
correlation on these roughly beam-sized angular scales does not
strongly influence the overall correlation statistics.  When the
statistics of the correlation functions are calculated excluding
correlations on angular scales smaller than $1\arcdeg$, the values of
$\chi^2_{\nu}$ show only modest decreases at best, and the
corresponding probabilities for isotropy are only slightly improved.
Examples of these are given in the last columns of Table 5.

Finally, to place limits on the {\it anisotropy} of the 240 $\micron$
residual emission within the HQB region, a technique commonly used to
limit temperature fluctuations in the CMB is employed (e.g., Readhead
et al. 1989; Church et al. 1997). The observed correlation function is
compared with Gaussian autocorrelation function (GACF) models of the
form $C(\theta)=C_0(\theta_c)\exp{(-\theta^2/ 2\theta_c^2)}$, where
$\theta_c$ is the intrinsic correlation scale of the fluctuations and
$C^{1/2}_0(\theta_c)$ is their mean amplitude. Convolution of
intrinsic fluctuations with a Gaussian approximation to the DIRBE beam
[$\exp(-\theta^2/2\theta_0^2)$ with $\theta_0\approx 0.3^\circ$] gives
a correlation function of the form:
$$
C(\theta)=C_0(\theta_c){{\theta_c^2}\over{2\theta_0^2+\theta_c^2}}
\exp{\biggl[-{{\theta^2}\over{2(2\theta_0^2+\theta_c^2)}}\biggr]}. \eqno(4)
$$
Fitting this model to the data provides limits on $C_0(\theta_c)$. For
the HQB region and the 240 $\micron$ residual emission after removal
of the ISM1 model, the best fit GACF has an amplitude of
$C_0(\theta_c){{\theta_c^2}\over{2\theta_0^2+\theta_c^2}}=10\pm2\nWm2sr$
and an apparent scale length $2\theta_0^2+\theta_c^2\approx2\theta_0^2$.
If this correlation is removed, there is no other correlation on
angular scales larger than $\sim 2\arcdeg$, limited by
$C_0(\theta_c)<1\nWm2sr$.  For the 240 $\micron$ residual emission
after removal of the ISM2 model, correlation is again found on a scale
comparable to the beam, but with an increased amplitude of
$C_0(\theta_c){{\theta_c^2}\over{2\theta_0^2+\theta_c^2}}=50\pm20
\nWm2sr$. Fluctuations on other scales can not be limited as tightly
as for the residual emission of the ISM1 model subtraction.  The
small-scale angular correlation appearing in the first several angular
bins of the 240 $\mu$m plot has been investigated.  Residual
structures in the IPD cloud and interstellar medium do not produce
effects this large.  The extrapolated emission of the sources in the
{\it IRAS} Point Source Catalog also does not produce this much
correlated power.  After the small-scale angular correlation was found
in the residual maps, a weak time correlation in successive samples of
the 240 $\mu$m dark noise data (DIRBE shutter closed) was found.  This
temporal correlation maps into adjacent pixels in the sky, and is
large enough to produce the observed small-angle correlation.
However, the cause of this unexpected instrumental effect is not
known.

\def\nodata{ ~$\cdots$~ }
\footnote{}
{\eightpoint\vskip6pt
 \tabbot{5}{R{\sc ESULTS FROM} T{\sc WO}-P{\sc OINT} C{\sc ORRELATION}}
 {
  #\hfil&\hfil#\hfil&\hfil#\hfil&\hfil#\hfil&\hfil#\hfil&
  \hfil#\hfil&\hfil#\hfil\cr
  \tabhead{
   \noalign{\hskip3.0in{$0\arcdeg\leq\theta\leq180\arcdeg$}
            \hskip1.5in{$1\arcdeg\leq\theta\leq180\arcdeg$}}
   \noalign{\vskip-3pt\hskip2.7in${\hskip1.3in\over\hskip1.3in}$
                      \hskip0.9in${\hskip1.3in\over\hskip1.3in}$
            \vskip-3pt}
   {Wavelength ($\micron$)}&{Location}&{$\chi^2_{\nu}$\tablenotemark{a}}&
   {$P(<\chi^2_{\nu})$\tablenotemark{b}}&{}&{$\chi^2_{\nu}$\tablenotemark{a}}&
   {$P(<\chi^2_{\nu})$\tablenotemark{b}}\cr}
  \tabbody{
140 & HQB  & 1.31 & 0.99 & & 1.24 & 0.97\nl
140 & HQBN & 1.24 & 0.92 & & 1.16 & 0.85\nl
140 & HQBS & 1.09 & 0.75 & & 1.05 & 0.67\nl
140\tablenotemark{c} & LH & 1.27 & 0.91 & & \nodata & \nodata \nl
240 & HQB  & 1.19 & 0.95 & & 1.13 & 0.87\nl
240 & HQBN & 1.34 & 0.97 & & 1.31 & 0.96\nl
240 & HQBS & 1.06 & 0.68 & & 0.98 & 0.48\nl
240\tablenotemark{c} & LH & 0.94 & 0.41 & & \nodata & \nodata \nl
240\tablenotemark{d} & HQB  & 1.10 & 0.83 & & 1.10 & 0.81\nl
240\tablenotemark{d} & HQBN & 0.99 & 0.52 & & 1.00 & 0.54\nl
240\tablenotemark{d} & HQBS & 1.13 & 0.80 & & 1.13 & 0.81\nl
\noalign{\vskip6pt\hrule\vskip6pt}
140 $\times$ IPD & HQB  & 1.21 & 1.00 & & \nodata & \nodata \nl
140 $\times$ IPD & HQBN & 1.09 & 0.81 & & \nodata & \nodata \nl
140 $\times$ IPD & HQBS & 1.30 & 1.00 & & \nodata & \nodata \nl
140 $\times$ ISM & HQB  & 1.28 & 1.00 & & \nodata & \nodata \nl
140 $\times$ ISM & HQBN & 1.16 & 0.93 & & \nodata & \nodata \nl
140 $\times$ ISM & HQBS & 1.15 & 0.92 & & \nodata & \nodata \nl
240 $\times$ IPD & HQB  & 1.12 & 0.95 & & \nodata & \nodata \nl
240 $\times$ IPD & HQBN & 0.99 & 0.48 & & \nodata & \nodata \nl
240 $\times$ IPD & HQBS & 1.17 & 0.95 & & \nodata & \nodata \nl
240 $\times$ ISM & HQB  & 1.13 & 0.95 & & \nodata & \nodata \nl
240 $\times$ ISM & HQBN & 1.03 & 0.63 & & \nodata & \nodata \nl
240 $\times$ ISM & HQBS & 1.07 & 0.77 & & \nodata & \nodata \nl
240\tablenotemark{d} $\times$ IPD & HQB  & 1.21 & 1.00 & &\nodata & \nodata \nl
240\tablenotemark{d} $\times$ IPD & HQBN & 1.05 & 0.71 & &\nodata & \nodata \nl
240\tablenotemark{d} $\times$ IPD & HQBS & 1.29 & 1.00 & &\nodata & \nodata \nl
240\tablenotemark{d} $\times$ ISM & HQB  & 1.16 & 0.98 & &\nodata & \nodata \nl
240\tablenotemark{d} $\times$ ISM & HQBN & 1.06 & 0.71 & &\nodata & \nodata \nl
240\tablenotemark{d} $\times$ ISM & HQBS & 1.06 & 0.73 & &\nodata & \nodata \nl
 }}
 \tabnote{$^a$ For $0\arcdeg\leq\theta\leq180\arcdeg$: $\nu=360$ for
               HQB and $\nu=180$ for HQBN and HQBS. For
               $1\arcdeg\leq\theta\leq 180\arcdeg$: $\nu=356$ for HQB
               and $\nu=176$ for HQBN and HQBS.}
 \vskip18pt\hsize=\h@size
 \tabnote{$^b$ The probability of one Monte Carlo simulation having a
               smaller $\chi^2_{\nu}$ than the value listed in the
               preceding column.}
 \vskip18pt\hsize=\h@size
 \tabnote{$^c$ Residuals after subtraction of the Snowden et
               al. (1994) H~I data as the ISM model.}
 \vskip18pt\hsize=\h@size
 \tabnote{$^d$ Residual at 240 $\micron$ after subtraction of the
              2-component ISM model (ISM2).}
\vskip6pt
}

\subsection{3.6. Conclusions from Residuals}

The signatures of a candidate CIB detection are a significantly
positive residual and isotropy over the tested area of the sky.  We
require that a significant mean residual exceed $3\sigma$, where the
uncertainty, $\sigma$, is the quadrature sum of the random errors and
systematic uncertainties of the measurements and foreground removal.
The smallest sky area considered meaningful for isotropy testing is
the 2\% of the sky where there are generally minimal foregrounds, the
HQB region.

Within the HQB region, there are gradients in the residual emission
and little or no consistency with isotropy in the two-point
correlation functions and other isotropy tests for all wavelengths
from 1.25 to 100 $\micron$.  Furthermore, at all of these wavelengths,
with the exception of 4.9 $\mu$m, the mean residual emission is less
than $3\sigma$.  Therefore, from 1.25 to 100 $\micron$, we are only
able to establish upper limits on an isotropic background.  Using the
HQB analysis, upper limits at the 95\% confidence level (CL) are taken
to be the residual intensities, $\nu I_0({\rm HQB})$, plus twice the
quadrature sum of their random and systematic uncertainties.  These
upper limits are listed as $\nu I_0$ (95\% CL) in Row 17 of Table 2.

At 140 and 240 $\mu$m the two-point correlation functions indicate
that the residual emission is isotropic over the HQB region,
particularly if the North and South halves of the region are
considered separately (Table 5).  The absence of significant gradients
with ecliptic or galactic latitude (Table 2, rows 15--16) also
supports this conclusion.  However, the mean residuals at these
wavelengths in the HQB region alone does not exceed 3$\sigma$,
primarily as a result of the large systematic uncertainty arising from
using the 100 $\mu$m map as the ISM template.  As discussed in \S~3.4,
direct correlation of the infrared emission with the H~I column
density in the well-studied LH$^\prime$ region at the Lockman hole
results in smaller systematic uncertainties in the residual
intensities than yielded by our map-based procedures for subtracting
the ISM contribution in HQB.  The same is true at 240 $\mu$m for the
NEP$^\prime$ region at the north ecliptic pole. In particular, the
correlation procedure yields residual intensities in the Lockman hole
region that are greater than $3\sigma$ at 100, 140 and 240 $\micron$,
and that are consistent with the mean residuals and their
uncertainties in the HQB and NEP$^\prime$ regions.

In order to make full use of the most accurate determinations of the
residuals at these long wavelengths, the weighted average of the
residuals in the HQB, LH$^\prime$(HI) and NEP$^\prime$(HI) regions was
determined.  The weighting factors are the inverse squares of the
combined random and non-common-mode systematic uncertainties for the
three regions.  For these purposes, gain, offset, and IPD model errors
were considered common-mode errors, leaving the uncertainty in the ISM
removal as the systematic error.  This weights the LH$^\prime$(HI)
determination most heavily.  Row 18 of Table 2 shows the resulting
weighted averages, $\langle\nu I_0\rangle$, at 100, 140, and 240
$\mu$m.  The uncertainties in these values include the formal
propagated uncertainty of the averaging process added in quadrature
with the common-mode systematic uncertainties excluded in the
averaging.

Since the weighted-average residuals at 140 and 240 $\micron$, $\nu
I_\nu=25\pm~7$ and $14\pm~3\nWm2sr$ respectively, exceed $3\sigma$ and
satisfy the isotropy tests, these residuals are either detections of
the CIB or unmodeled isotropic contributions from sources in the solar
system or Galaxy.  Arguments against the foreground interpretation are
presented in Paper IV and summarized in \S~4.1.  Though the
weighted-average residual at 100 $\micron$ in the HQB region, Lockman
hole, and north ecliptic pole regions exceeds $3\sigma$, the
anisotropy in the HQB region excludes this as a candidate detection of
the CIB.  The anisotropy at $100~\mu$m may be the result of inaccuracy
in the ISM model due to use of the H~I template (\S~3.4), as well as
the appreciable artifacts from the IPD model at this wavelength.  The
weighted-average residual at $100~\mu$m provides a slightly more
restrictive upper limit on the CIB than the HQB analysis alone.  The
last row of Table 2 shows the values of the two measurements of the
CIB at 140 and 240 $\mu$m and the most restrictive upper limits at all
other wavelengths.

\section{4. DISCUSSION AND CONCLUSIONS}

\subsection{4.1. Possible Contributions from}
\vskip-12pt
\subsection{Unmodeled Isotropic Sources}

In order to verify that the probable isotropic residual emission at
140 and 240 $\micron$ is of extragalactic origin, we need to
demonstrate that local contributions of isotropic or nearly isotropic
components, both within the solar system and within the Galaxy, do not
contribute significantly to the residual emission.  Circumterrestrial
material is ruled out by lack of variation of the measured sky
brightness with zenith angle, and by the low color temperature of the
residual radiation.  Heliocentric material within the solar system may
have escaped our modeling efforts if it lies in the outer solar
system, where its intensity will show little or no modulation as the
Earth moves along its orbit. Such a cloud would not have been
detectable by the IPD modeling procedures applied, which relied on the
apparent temporal variations of the IPD emission. An isotropic
component of the Galactic emission may not have been removed by our
models if it arises from sources distributed in a roughly spherical
halo around the Galactic center of radius much larger than 8.5 kpc.

These potential solar system and Galactic sources are considered in
detail in Paper IV.  In the case of the solar system, it is shown that
a spherical cloud formed early in the history of the solar system
would not survive to the present.  A persistent spherical cloud would
require a source of replenishment, and no plausible source for a cloud
of adequate mass can be identified.  Difficulties with attributing a
significant portion of the 140 and 240 $\micron$ isotropic residual
emission to a Galactic dust component include the lack of a plausible
mechanism for creating and maintaining a large, smooth, shell-like
distribution of dust, and the absence of a heating source which could
maintain a uniform dust temperature as high as that implied by the
detections ($\sim$17 K) at large distances from the Galactic plane.
Furthermore, such a shell would require such a large dust mass that
the associated gas mass would be at least comparable to that in the
Galactic disk (assuming metallicity no greater than solar).

Hence, there is no known or likely source, consistent with other
present knowledge of the solar system and Galaxy, which can meet the
combination of constraints imposed by the low color temperature and
isotropy of the long wavelength residual detections.  We conclude that
it is unlikely that significant fractions of the observed 140 and 240
$\micron$ residual emission can arise from either an IPD or a Galactic
emission component.  The most likely conclusion is that these signals
arise from an extragalactic infrared background.

\subsection{4.2 Comparison with Previous Limits}

\subsection{4.2.1. {\it Direct Infrared Brightness Measurements}}

Figure~4 summarizes the current state of direct infrared background
measurements.  DIRBE results presented in this paper are shown from
1.25 to 240 $\mu$m for both the dark sky upper limits ($2\sigma$ above
the lowest measured values, from Table 2) and the limits and
detections after foreground removal.  Dark sky upper limits from 120
to 650 $\mu$m determined in ``broad bands" from {\it COBE}/FIRAS data
(Shafer \etal 1998) are also shown.  In the 140--240 $\mu$m region,
the FIRAS dark sky limits are in excellent agreement with the
corresponding DIRBE limits.  Since the calibrations of the two
instruments are very consistent (Fixsen \etal 1997), this suggests
that there are no small regions (on the scale of the DIRBE beam) in
which the DIRBE has a better view beyond the Galaxy than does the
FIRAS with its much larger beam.

Near-infrared limits from recent rocket measurements (Matsuura \etal
1994) are similar to the DIRBE dark sky limits, whereas the ``unknown
residual emission'' after foreground removal by Noda \etal (1992) is
close to the foreground-removed DIRBE upper limits.  For comparison,
an upper limit obtained from sky photometry in the optical is also
shown (Mattila 1990).  Far-infrared limits from the rocket data of
Kawada \etal (1994) are generally similar to the {\it COBE} dark sky
values, though the quoted residual upper limit at 154 $\mu$m is very
close to the DIRBE detection at 140 $\mu$m.  Schlegel, Finkbeiner, \&
Davis (1997) have recently studied Galactic reddening using DIRBE long
wavelength data as a tracer of the interstellar dust.  Using a simpler
long-wavelength IPD model than ours (Paper II), they have found
uniform backgrounds at 140 and 240~$\mu$m which they identify as CIB
detections at levels similar to the residual values reported here.

Figure~4 also shows the tentative detection of a $170-1260~\mu$m
background based upon FIRAS data reported by Puget \etal (1996).
This result is significantly below the $140-240~\mu$m detections
reported here.  Even if the DIRBE result were to be recalibrated using
the DIRBE--FIRAS calibration comparison of Fixsen et al. (1997), this
significant difference would remain.  We have no ready explanation for
that difference.  However, Fixsen et al. (1998) have recently
completed an extensive assessment of the evidence for the CIB in the
FIRAS data.  In order to investigate the magnitude of the systematic
uncertainties involved in separating Galactic emission from the CIB,
they have used three independent methods to derive the CIB spectrum.
One of these methods assumes that our DIRBE results are correct, and
so we ignore that one here for purposes of comparing the DIRBE and
FIRAS results.  Figure~4 shows the average of the Fixsen et al. (1998)
results using two other methods for separation of Galactic emission: a
method based upon assuming a single color temperature for the ISM
emission; and a method using maps of H~I and C~II emission to trace
the ISM.  Convolving this average of the results of Fixsen et
al. (1998) with the DIRBE spectral responses at 140 and 240~$\mu$m
yields FIRAS values at the same effective wavelengths of $\nu
I_\nu=11.5$ and $11.3\nWm2sr$ respectively.  These values are within
$2\sigma$ and $1\sigma$ of the DIRBE results (Table 2) respectively,
and so are entirely consistent with them.  If we formally transform
the DIRBE results to the FIRAS photometric scale according to the
determination of Fixsen et al. (1997), we obtain $\nu I_\nu=15.0$ and
$12.7\nWm2sr$ at 140 and 240 $\mu$m respectively.  Thus, even the
small difference between the DIRBE 240 $\mu$m result and that of
Fixsen et al. (1998) arises in large part from the small difference in
photometric scales of the two instruments, and not in the separation
of the foreground radiations from the CIB.  The difference between the
experiments at 140 $\mu$m mostly arises from the calibration
difference.  We conclude that the FIRAS analysis of Fixsen et
al. (1998) provides strong independent confirmation of the DIRBE
observational conclusions.

{\topinsert\null\vskip-12pt
  \psfig{file=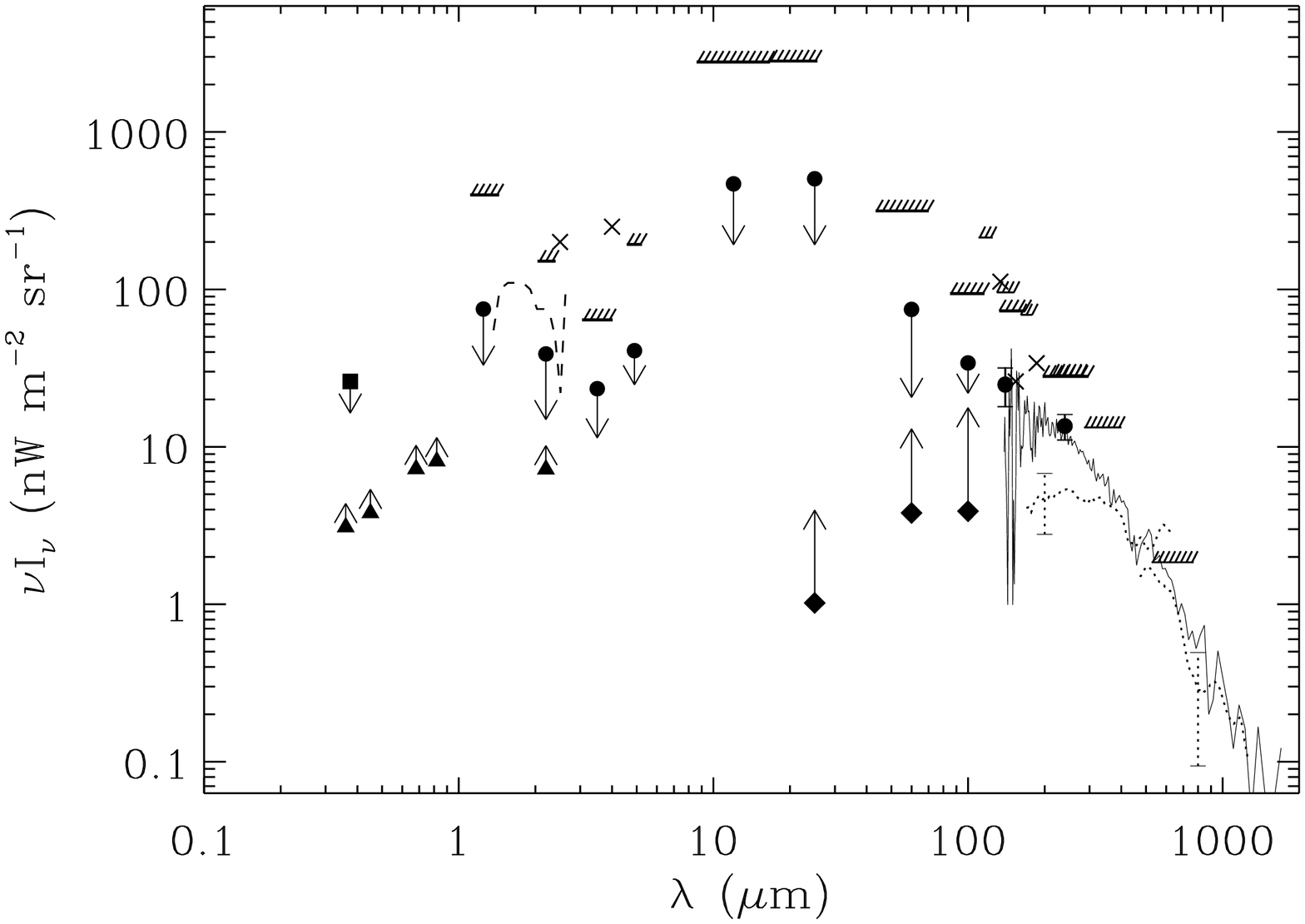,width=\h@size,angle=90,silent=yes}
  \vskip12pt\edef\h@old{\the\hsize}\hsize=\h@size
  Fig.~{4}.---{Cosmic background intensity $I_\nu$ times frequency
  $\nu$ as a function of wavelength $\lambda$. The circles with error
  bars are the detections based on DIRBE data after removal of
  foreground emission at 140 and $240~\mu$m, while those with arrows
  are 2$\sigma$ upper limits with the arrows extending to the measured
  residuals at $1.25-100~\mu$m.  The hatched thick lines are dark sky
  limits (95\% CL) from the DIRBE data at $1.25-240~\mu$m, while the
  hatched thin lines are dark sky ``broad--band" limits (95\% CL) from
  FIRAS data at $120-650~\mu$m (Shafer et al. 1998). The crosses are
  upper limits derived from rocket experiments at $134-186~\mu$m
  (Kawada et al. 1994) and $2.5-4.0~\mu$m (Matsuura et al. 1994). The
  dashed line from $1.4-2.6~\mu$m is residual radiation after
  foreground removal from the rocket data of Noda \etal (1992).  The
  diamonds with arrows are lower limits derived from {\it IRAS} counts
  at $25-100~\mu$m (Hacking \& Soifer 1991; $60~\mu$m limit from
  Gregorich et al. 1995). The dotted curve from $170-1260~\mu$m shows
  the tentative infrared background determined from FIRAS data by
  Puget et al. (1996), while the solid curve is the average of the two
  DIRBE-independent methods of FIRAS analysis used by Fixsen et
  al. (1998).  The triangles are lower limits derived from the Hubble
  Deep Field at $3600-8100$~{\AA} (Pozzetti et al. 1998) and $K$-band
  galaxy counts at $2.2~\mu$m (Cowie et al. 1994). The square is an
  upper limit derived from sky photometry at $4400$~{\AA} (Mattila
  1990).}
  \vskip24pt\vskip-5.23433pt
  \hsize=\h@old\endinsert\vfuzz=0.5pt}

\subsection{4.2.2. {\it Angular Fluctuation Limits}}

An alternative approach to searching for evidence of the CIB is to
study the fluctuations in maps of the infrared sky brightness.  If the
spatial correlation function of the sources is known, the diffuse
background produced by them can be estimated from the measured
correlation function of sky brightness.  Using such arguments,
Kashlinsky \etal (1996a) obtained upper limits on the CIB from
clustered matter of 200, 78, and $26~\nWm2sr$ at 1.25, 2.2, and 3.5
$\mu$m respectively, values modestly above the present direct DIRBE
brightness limits in Table~2 and Figure~4.  In an extension of this
approach, Kashlinsky \etal (1996b) determined the rms fluctuations in
the DIRBE maps from $2.2-100~\mu$m, and argue that these values imply
that the CIB due to matter clustered like galaxies is less than about
$10-15~\nWm2sr$ over this wavelength range.  In the near infrared and
at 100~$\mu$m, these values are close to the observed residuals
reported in Table 2.  In the thermal infrared region, $12-60~\mu$m,
where the accurate removal of the large contribution from the
interplanetary dust is so difficult, these limits are much below the
limits reported here.  However, relating the limit on rms map
fluctuations to the absolute brightness of the sky does involve
model-dependent assumptions about the clustered sources of radiation.

\subsection{4.2.3. {\it Limits from TeV Gamma Rays}}

Indirect evidence for the CIB can be obtained in principle by
observing attenuation of very energetic $\gamma$-rays from
extragalactic sources (Gould \& Schreder 1967).  Attenuation will
arise from pair-production in the interaction of the $\gamma$-rays
with infrared photons.  Such arguments, based upon apparent evidence
for attenuation of TeV $\gamma$-rays from Mk 421, have been used to
obtain both upper and lower limits on the CIB.  The limits obtained
depend on the assumed spectrum of the CIB, as well as of the intrinsic
spectrum of the $\gamma$-ray source (de Jager, Stecker, \& Salamon
1994; Dwek \& Slavin 1994; Biller et al. 1995; Stecker 1996; Stecker
\& de Jager 1997).  However, Krennrich \etal (1997) have recently
reported detection of $\gamma$-rays with energies exceeding 5 TeV from
Mk 421.  These authors conclude that there is no present evidence in
the data for attenuation by pair production on optical or
near-infrared photons, though given the uncertainty in the intrinsic
$\gamma$-ray source spectrum, the possibility of some such attenuation
can not be totally ruled out.  Under the above assumptions, even with
no evident attenuation, these observations provide upper limits on the
CIB between 15 and 40 $\mu$m of about $10-20~\nWm2sr$ (e.g., Dwek \&
Slavin 1994).  These limits are well below the present direct limits
from DIRBE data, and are comparable to those obtained by Kashlinsky,
Mather \& Odenwald (1996b) from their analysis of fluctuations in the
DIRBE maps (\S~4.2.2).  Recent analysis of the TeV $\gamma$-ray data
from Mk 501 by Stanev and Franceschini (1997) yields limits from 1 to
40 $\mu$m in the range $1-20~\nWm2sr$ depending upon the assumed
spectrum of the CIB.  Because of the large observational and
theoretical uncertainties inherent in these limits, we do not yet
regard them as strong constraints on currently popular theoretical
models of the CIB in this wavelength interval (Paper IV).

\subsection{4.3. Relationship to Integrated Brightness of Galaxies}

Lower limits to the extragalactic infrared background can be obtained
by integrating the brightness of observed galaxies.  Figure 4 shows
such results from the near-infrared galaxy counts of Cowie \etal
(1994), and from the {\it IRAS} survey by Hacking \& Soifer (1991) and
Gregorich \etal (1995).  The {\it IRAS} results are shown as a range
to encompass the various galaxy luminosity or density evolution models
considered. Figure~4 also shows lower limits at UV and optical
wavelengths derived from galaxy counts in the Hubble Deep Field
(Pozzetti \etal 1998).  It is comforting to see that the integrated
discrete source estimates still lie below the diffuse sky brightness
residuals, and the gap is not large at some wavelengths.  For example,
the bright end of the evolution models considered by Hacking \& Soifer
(1991) at 60 and 100 $\mu$m [as amended by Gregorich \etal (1995) at
60 $\mu$m] is only about a factor of two below the DIRBE measured
residuals at the corresponding wavelengths.  The estimated integrated
galaxy far-infrared background contribution should become less
uncertain as deeper counts from space missions such as {\it ISO}, {\it
WIRE}, and {\it SIRTF} are obtained.

\subsection{4.4. Limit on Integrated Infrared Background}

The CIB limits and detections reported here provide an upper limit on
the integrated energy density of the CIB, an overall constraint on the
integrated cosmic luminosity.  Denoting the integrated infrared
background energy density in units of the critical closure energy
density by $\Omega_{IR}$ and the corresponding quantity for the CMB by
$\Omega_{CMB}$, one finds that (for $T_{CMB}$ = 2.728 K, Fixsen et al.
1996) $\Omega_{IR}/\Omega_{CMB}=1\times10^{-3}\times(I_{IR}/\nWm2sr$),
where $I_{IR}$ is the sky brightness integrated over the infrared
spectrum.  Taking the range of integration for the infrared to be $1-
300~\mu$m, the dark sky upper limits of Table 2 give $\Omega_{IR}/
\Omega_{CMB}<2.4$, not a very restrictive limit.  If the DIRBE upper
limits plus likely detections shown in Table 2 are used, one finds an
upper limit of $\Omega_{IR}/\Omega_{CMB}<0.5$.

To provide substantially more stringent limits on the integrated
infrared background over this broad spectral range, the peak in the
limits over the thermal infrared range ($\sim5-60~\mu$m), which may
largely be due to the difficulty in discriminating the IPD signal to
better than a few percent of its value, must be substantially reduced.
However, the limits on both the short-wavelength and long-wavelength
sides of this peak are themselves of interest, since they constrain
both the directly radiated energy density and that due to primary
radiation absorbed by dust and re-emitted at longer wavelengths.  The
strong upper limits found from the dark sky upper limits of Table 2
are $\Omega_{IR}/\Omega_{CMB}< 0.16$ and $\Omega_{IR}/\Omega_{CMB}<
0.05$ in the ranges $1-5~\mu$m and $100-240~\mu$m respectively.  Using
the foreground-removed upper limits and detections from Table 2, the
corresponding limits are $\Omega_{IR}/\Omega_{CMB}<0.04$ and
$\Omega_{IR}/\Omega_{CMB}<0.02$.

\subsection{4.5. Implications}

The DIRBE CIB detections and upper limits cover a broad spectral range
from $1.25~\mu$m to $240~\mu$m.  The CIB intensity in the
$1.25-5~\mu$m range is likely dominated by direct starlight from
galaxies, whereas the intensity in the $100-240~\mu$m range is likely
dominated by reradiated starlight from dust within galaxies.  Under
these assumptions, one of the important implications of the DIRBE
results is that they provide valuable constraints on the global
history of star formation and dust production in the universe.  In
general, the CIB is a fossil containing the cumulative energy release
of astrophysical objects or processes in the universe.  The DIRBE
results can therefore be used to discriminate and constrain possible
contributors to the CIB, such as active galactic nuclei, halo black
holes, pregalactic stars, decaying particles, and gravitational
collapse (e.g., Bond, Carr, \& Hogan 1991).  Here we briefly discuss
the implications of our measurements for star formation and dust
production in galaxies based largely upon published models.  Paper IV
provides more extensive discussion of the cosmological implications.

One of the surprising consequences of the DIRBE results presented here
is that the detected energy level of the far-IR background,
$\int\nu I_\nu d\ln\nu=10.3~ {\rm nW~m}^{-2}~ {\rm sr}^{-1}$
in the $140-240~\mu$m range, is a
factor of $\sim 2.5$ higher than the integrated optical light from the
galaxies in the Hubble Deep Field, $\int\nu I_\nu d\ln\nu=4.2\nWm2sr$
in the $3600-8100$~\AA\ range (Pozzetti et al. 1998).  Since the full
spectrum of the cosmic background in the UV-optical and far-infrared
wavelength ranges is unknown, the exact ratio of the backgrounds in
these ranges is still quite uncertain.  Nevertheless, the DIRBE
detections, when compared with the Hubble Deep Field results, indicate
that a substantial fraction of the total stellar luminosity from
galaxies might have been reradiated by dust in the far-infrared at the
expense of the obscured UV-optical luminosity.  This implies that star
formation might be heavily shrouded by dust at high redshifts.

Figure~5 shows the same data as in Figure~4, superimposed on CIB
estimates for some early models of possible pregalactic and
protogalactic sources in a dust-free universe (Bond, Carr, \& Hogan
1991).  Clearly the DIRBE upper limits in the near-infrared and the
lower limits from deep optical and near-infrared galaxy counts either
rule out such models or require revision of their parameters.
Figure~5 also shows two examples of the predicted contributions of
galaxies to the CIB.  The dashed curves are the calculations of
Franceschini et al.  (1994) using evolutionary models with moderate
and opaque dust optical depth, largely based on emission properties of
galaxies at the present epoch.  The solid curves are the calculations
of Fall, Charlot, \& Pei (1996) using closed-box and inflow models of
cosmic chemical evolution, largely based on absorption properties of
galaxies at different redshifts.  Both classes of models modestly
underpredict the DIRBE measurements of the far-infrared background.
Since star formation and dust production are coupled, fitting CIB
estimates from models of cosmic chemical evolution to the DIRBE
detections can determine the rates of both star formation and dust
production as a function of redshift.  In this fashion, the DIRBE
results taken together with deep optical surveys of galaxies promise
to yield improved estimates of the history of global star formation,
metal and dust production, and the efficiency of UV-optical absorption
by dust.

{\topinsert\null\vskip-12pt
  \psfig{file=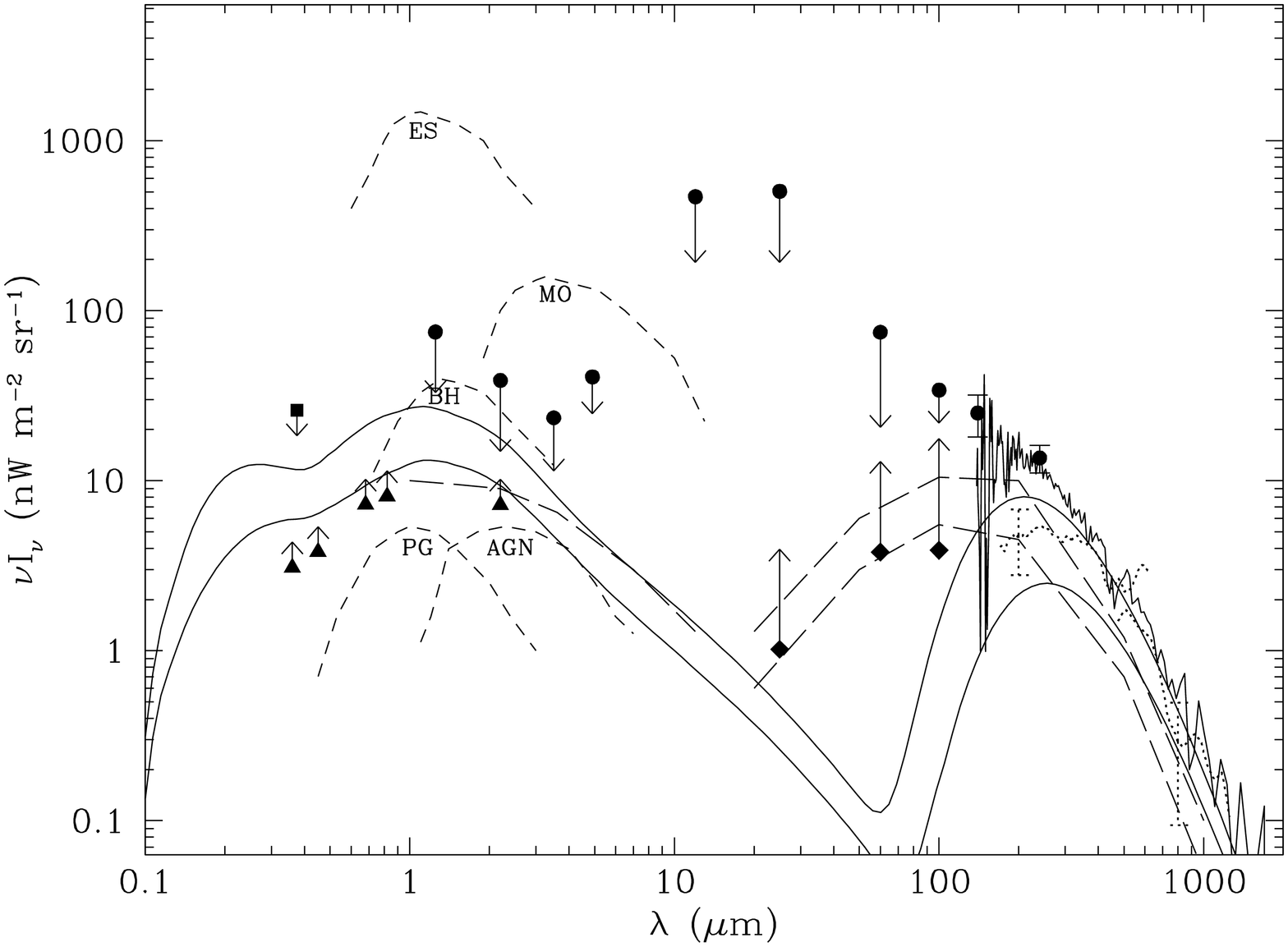,width=\h@size,angle=270,silent=yes}
  \vskip12pt\edef\h@old{\the\hsize}\hsize=\h@size
  Fig.~{5}.---{Predicted contributions to the cosmic infrared
  background radiation.  The data points and FIRAS curves are the
  same measurement results as in Fig.~4.  The short-dashed lines show
  CIB estimates by Bond, Carr, \& Hogan (1991) for some possible
  pregalactic and protogalactic sources in a dust-free universe,
  including exploding stars (ES), massive objects (MO), halo black
  holes (BH), active galactic nuclei (AGN), and primeval galaxies (PG)
  .  The long-dashed lines are calculations of Franceschini et
  al. (1994) using models of photometric evolution of galaxies with
  two cases of dust opacity.  The solid lines are calculations of
  Fall, Charlot, \& Pei (1996) using closed-box (lower curve) and 
  inflow (upper curve) models of  cosmic chemical evolution.}
  \vskip24pt\vskip-5.23433pt
  \hsize=\h@old\endinsert\vfuzz=0.5pt}

\subsection{4.6. Summary}

The DIRBE investigation was designed to detect directly the CIB, or
set limits on it imposed by the brightness of our local cosmic
environment.  The observational results reported here, supported by
Papers II and III, show evidence for detection of such a background at
the level of $25\pm7\nWm2sr$ at 140 $\micron$ and $14\pm3\nWm2sr$ at
240 $\micron$, and upper limits at wavelengths from 1.25 to 100
$\micron$.  As our analyses show, the uncertainties in these results
are indeed dominated by the uncertainties in our ability to
discriminate or model the contributions to the infrared sky brightness
from sources within the solar system and Milky Way Galaxy.

These results very substantially advance our prior direct knowledge of
the extragalactic infrared sky brightness, especially of what was
known prior to the {\it COBE} mission.  The quality of the DIRBE
measurements themselves is such that improved knowledge of the local
foregrounds could permit the search for the CIB to be carried to more
sensitive levels using DIRBE data. Such knowledge will be provided by
future measurements, such as the sensitive all-sky surveys at 2
microns (2MASS and DENIS) and more extensive measurements of Galactic
H~II emission at high latitudes, and possibly by improved techniques
to model or discriminate the very dominant contribution from
interplanetary dust (e.g., Gor'kavyi et al. 1997a; 1997b).  Of course,
further direct measurements of the absolute infrared sky brightness
with higher angular resolution, preferably from a location more
distant from the Sun so as to reduce the contribution of the
interplanetary dust to the sky brightness, could advance this search
dramatically.

\vskip34pt

The authors gratefully acknowledge the contributions over many years
of the talented and dedicated engineers, managers, scientists,
analysts, \hskip10pt and programmers \hskip10pt engaged in the DIRBE
investigation.  We thank G. Hinshaw for expert advice on two-point
correlation functions.  The National Aeronautics and Space
Administration/Goddard Space Flight Center (NASA/GSFC) was responsible
for the design, development, and operation of the {\it COBE}.
Scientific guidance was provided by the {\it COBE} Science Working
Group.  GSFC was also responsible for the development of the analysis
software and for the production of the mission data sets.

\vskip36pt
\centerline {REFERENCES}
\vskip12pt

\def\apj{ApJ}\def\aj{AJ}\def\apjs{ApJS}\def\procspie{Proc. SPIE}
\def\aaps{A\&AS}\def\nat{Nature}
\def\reference{\ref}

\reference{} Arendt, R. G., Odegard, N., Weiland, J. L., Sodroski, T. J.,
Hauser, M. G., Dwek, E., Kelsall, T., Moseley, S. H., Silverberg, R. F.,
Leisawitz, D., Mitchell, K., \& Reach, W. T. 1998, ApJ, submitted (Paper III)
  
\reference{} Biller, S. D., \etal 1995, \apj, 445, 227

\reference{} Boggess, N. \etal 1992, \apj,  397, 420 

\reference{} Bond, J. R., Carr, B. J., \& Hogan, C. J. 1986, \apj,
306, 428

\reference{} Bond, J. R., Carr, B. J., \& Hogan, C. J. 1991, \apj,
367, 420

\reference{} Cohen, M. 1993, \aj, 105, 1860

\reference{} Cohen, M. 1994, \aj, 107, 582

\reference{} Cohen, M. 1995, \apj, 444, 875

\reference{} Church, S. E., Ganga, K. M., Ade, P. A. R., Holzapfel,
W. L., Mauskopf, P. D., Wilbanks, T. M., \& Lange, A. E. 1997, \apj,
484, 523

\def\s{\hskip10pt}
\reference{} COBE\s Diffuse\s Infrared\s Background\s Experiment (DIRBE)
Explanatory Supplement, version 2.1, ed. M. G. Hauser, T. Kelsall, D.
Leisawitz, and J. Weiland, COBE Ref. Pub. No.  97-A (Greenbelt, MD:
NASA/GSFC), available in electronic form from the NSSDC at {\tt
http://www.gsfc.nasa.gov/astro/cobe /cobe\_home.html}

\reference{} Cowie,~L.~L., Gardner,~J.~P., Hu,~E.~M., Songaila,~A.,
Hodapp,~K.-W., \& Wainscoat,~R.~J. 1994, ApJ, 434, 114

\reference{} de Jager, O. C., Stecker, F. W. \& Salamon, M. H.  1994,
\nat, 369, 294

\reference{} Dwek, E. \& Slavin, J. 1994, \apj, 436, 696

\reference{} Dwek, E., Arendt, R. G., Hauser, M. G., Fixsen, D.,
Leisawitz, D., Pei, Y.C., Wright, E. L., Kelsall, T., Mather, J. C.,
Moseley, S. H., Odegard, N., Shafer, R., Silverberg, R. F., \&
Weiland, J. L. 1998, ApJ, submitted (Paper IV)

\reference{} Elvis, M., Lockman, F.J., \& Fassnacht, C. 1994, \apjs,
95, 413.

\reference{} Evans, D. C. 1983, \procspie, 384, 82

\reference{} Fall, S.M., Charlot, S., \& Pei, Y.C. 1996, \apj, 464, L43

\reference{} Fixsen, D. J., Cheng, E. S., Gales, J. M., Mather, J. C.,
Shafer, R. A., \& Wright, E. L. 1996, \apj, 473, 576 

\reference{} Fixsen, D. J., Weiland, J. L., Brodd, S., Hauser, M. G.,
Kelsall, T., Leisawitz, D. T., Mather, J. C., Jensen, K. A., Shafer,
R. A., \& Silverberg, R. F. 1997, \apj, 490, 482

\reference{} Fixsen, D. J. \etal 1998, in preparation

\reference{} Franceschini, A., Toffolatti, L., Mazzei, P., \& de Zotti, G.
1991, \aaps, 89, 285

\reference{} Franceschini, A., Mazzei, P., de Zotti, G., \& Danese, L.
1994, \apj, 427, 140

\reference{} Gor'kavyi,~N.~N., Ozernoy,~L.~M., \& Mather,~J.~C.
1997a, \apj, 474, 496

\reference{} Gor'kavyi,~N.~N., Ozernoy,~L.~M., Mather,~J.~C., \&
Taidakova,~T. 1997b, \apj, 488, 268

\reference{} Gould,~R.~J. \& Schreder,~G. 1967, Phys. Rev., 155, 1408

\reference{} Gregorich,~D.~T., Neugebauer,~G., Soifer,~B.~T.,
Gunn,~J.~E., \& Herter,~T.~L. 1995, AJ, 110, 259
  
\reference{} Hacking,~P.~B., \& Soifer,~B.~T. 1991, ApJ, 367, L49

\reference{} Hartmann, D., \& Burton, W. B. 1997, Atlas of Galactic
Neutral Hydrogen, (New York: Cambridge University Press)

\reference{} Harwit, M. 1970, Rivista del Nuovo Cimento,  II, 253 

\reference{} Hauser, M. G. 1995, in Extragalactic Background
Radiation, Space Telescope Sci. Inst. Symp. Ser. 7, ed. D.  Calzetti,
M. Livio, \& P. Madau, (Cambridge: Cambridge Univ.  Press), 135

\reference{} Hauser, M. G. 1996a, in Proc. IAU Symposium 168, 
Examining the Big Bang and Diffuse Background Radiations, ed. M.
Kafatos \& Y. Kondo, (Dordrecht: Kluwer), 99

\reference{} Hauser,~M.~G. 1996b, in Unveiling the Cosmic Infrared
Background, ed. E.~Dwek (New York: IAP Press), 11

\reference{} Hinshaw, G., Banday, A.~J., Bennett, C.~L., G\'orski,
K.~M., Kogut, A., Lineweaver, C.~H., Smoot, G.~F., \& Wright, E.L.,
1996, \apj, 464, L25

\reference{} Jahoda, K., Lockman, F. J., \& McCammon, D. 1990, \apj,
354, 184

\reference{} Kashlinsky, A., Mather, J. C., Odenwald, S., \& Hauser,
M. G. 1996a, \apj, 470, 681

\reference{} Kashlinsky, A., Mather, J. C., \& Odenwald,~S.  1996b,
\apj, 473, L9

\def\s{\hskip18pt}
\reference{} Kawada,~M., Bock,~J.~J., Christov,~V.~V., Lange,~A.~E.,
Matsuhara,~H.,\s Matsumoto,~T.,\s Matsuura,~S., Mauskopf,~P.~D.,
Richards,~P.~L., \& Tanaka,~T. 1994, \apj, 425, L89

\reference{} Kelsall, T., Weiland, J. L., Franz, B. A., Reach, W. T.,
Arendt, R. G., Dwek, E., Freudenreich, H. T., Hauser, M. G., Moseley, S. H.,
Odegard, N. P., \& Silverberg, R. F. 1998, ApJ, submitted (Paper II)

\reference{} Krennrich,~F. \etal 1997, \apj, 481, 758

\reference{Lei96} Leinert, C. 1996, in Unveiling the Cosmic Infrared
Background, ed. E. Dwek, (Woodbury, NY: AIP), 53

\reference{} Lockman, F. J., Jahoda, K., \& McCammon, D. 1986, \apj,
302, 432

\reference{} Magner, T. J. 1987, OptEng,  26, 264 

\reference{} Matsumoto, T. 1990, in The Galactic and Extragalactic
Background Radiation, IAU Symposium 139, ed. S.  Bowyer \& C. Leinert
(Dordrecht: Kluwer), 317

\reference{} Matsumoto, T., Akiba, M., \& Murakami, H. 1988, \apj,
332, 575

\reference{} Matsuura,~S., Kawada,~M., Matsuhara,~T., Noda,~M., \&
Tanaka,~M. 1994, PASP, 106, 770

  
\reference{} Mattila,~K. 1990, in IAU Symposium 139, The Galactic and
Extragalactic Background Radiation, ed. S.~Bowyer \& C.~Leinert
(Dordrecht: Kluwer), 257

\reference{} Noda, M., Christov, V. V., Matsuhara, H., Matsumoto, T.,
Matsuura, S., Noguchi, K., \& Sato, S. 1992, \apj, 391, 456

\reference{} Partridge, R. B. \& Peebles, P. J. E. 1967, \apj, 148,
377

\reference{} Pozzetti, L., Madau, P., Zamorani, G., Ferguson, 
H. C., \& Bruzual, G. A. 1998, MNRAS, submitted

\reference{} Puget,~J.-L., Abergel,~A., Bernard,~J.-P., Boulanger,~F.,
Burton,~W.~B., D\'esert,~F.-X., \& Hartmann,~D.  1996, A\&A, 308, L5
  
\reference{} Readhead, A. C. S., Lawrence, C. R., Myers, S. T.,
Sargent, W. L.  W., Hardebeck, H. E., \& Moffet, A. T. 1989, \apj,
346, 566

\reference{} Ressell, M. T. \& Turner, M. S. 1990, Comments on
Astrophysics, 14, 323

\reference{} Schlegel, D. J., Finkbeiner, D. P., \& Davis, M. 1997,
preprint

\def\s{\hskip0.2pt}
\reference{} Shafer,~R.~A.,\s Mather,~J.~C.,\s Fixsen,~D.~J.,\s
Jensen,~K.~A., Reach,~W.~T., Wright,~E.~L., Dwek,~E., \&
Cheng,~E.~S. 1998, preprint


\reference{} Silverberg, R. F. \etal 1993, Proc. SPIE Conf. 2019, 
Infrared Spaceborne Remote Sensing, ed. M. S. Scholl (Bellingham:
SPIE), 180

\reference{} Snowden, S.L., Hasinger, G., Jahoda, K., Lockman, F.J.,
McCammon, D., \& Sanders, W.T. 1994, \apj, 430, 601

\reference{} Stanev, T. \& Franceschini, A. 1997, \apj, submitted

\reference{} Stark, A. A., Gammie, C. F., Wilson, R. W., Bally, J.,
Linke, R.  A., Heiles, C., \& Hurwitz, M. 1992, \apjs, 79, 77

\reference{} Stecker, F. W. 1996, in Unveiling the Cosmic Infrared
Background, ed. E. Dwek (Woodbury, NY: AIP), 181

\reference{} Stecker, F. W., \& de Jager, O. C. 1997, \apj, 476, 712

\reference{} Wainscoat, R. J., Cohen, M., Volk, K., Walker, H. J., \&
Schwartz, D. E. 1992, \apjs, 83, 111

\reference{} Wheelock, S.L. \etal 1994, IRAS Sky Survey Atlas
Explanatory Supplement, JPL Publication 94-11 (Pasadena:JPL)

\bye